%% file: main.tex
 \documentclass[sigconf]{acmart} 

\usepackage{multirow}
\usepackage{subcaption}
\usepackage{listings}
\usepackage{soul}
\usepackage{csquotes}
\usepackage{CJKutf8}
\usepackage{tikz}
\usepackage{tabularx}
\usepackage{longtable}
\usepackage{pdflscape}
\usepackage{pdfpages}
\usepackage{makecell}
\definecolor{agreegreen}{HTML}{2E7D32}     
\definecolor{agreered}{HTML}{C62828}       
\definecolor{disagreered}{HTML}{FAD4D4}    
\definecolor{mismatchamber}{HTML}{FFF2B2}  

\usepackage{arydshln}
\usepackage{pifont}
\usepackage{circledsteps}
\usepackage{placeins}
\definecolor{charcoalblue4F5A66}{HTML}{4F5A66}

\pgfkeys{
  /csteps/inner color=white,              
  /csteps/fill color=charcoalblue4F5A66,  
  /csteps/inner xsep=3pt,                 
  /csteps/inner ysep=2pt                
}
\makeatletter
\@ifundefined{Circled}{
\def\Circled#1{#1}
  }{}
\makeatother

\newcolumntype{P}[1]{>{\raggedright\arraybackslash}p{#1}}
\newcolumntype{L}{>{\raggedright\arraybackslash}T}

\definecolor{eclipseStrings}{RGB}{42,0.0,255}
\definecolor{eclipseKeywords}{RGB}{127,0,85}
\colorlet{numb}{magenta!60!black}

\newif\ifredact
\redacttrue   

\newif\ifcomment
\commenttrue   
\ifcomment
  \newcommand{\missing}[2][]{\textcolor{red}{[\textbf{MISSING\ifx#1\empty\else~–~#1\fi}] ~#2}}
  \newcommand{\kel}[1]{~\sethlcolor{pink}\hl{[Kellie: #1]}}
  \newcommand{\ken}[1]{~\sethlcolor{yellow}\hl{[Kenny: #1]}}
  \newcommand{\pinsym}[1]{~\sethlcolor{cyan}\hl{[Pin Sym: #1]}}
  \newcommand{\mel}[1]{~\sethlcolor{lightgray}\hl{[Melanie: #1]}}
  \newcommand{\chenyu}[1]{~\sethlcolor{green}\hl{[Chenyu: #1}}
\else
  \newcommand{\missing}[2]{}
  \newcommand{\kel}[1]{}
  \newcommand{\ken}[1]{}
  \newcommand{\pinsym}[1]{}
  \newcommand{\mel}[1]{}
  \newcommand{\chenyu}[1]{}
\fi

\newcommand{\acpagent}[1]{\textsc{ACPAgent}}
\newcommand{\baritone}[1]{\textbf{\textsf{P1}}}
\newcommand{\redpasta}[1]{\textbf{\textsf{P2}}}
\newcommand{\trombone}[1]{\textbf{\textsf{P3}}}
\newcommand{\cowalive}[1]{\textbf{\textsf{P4}}}
\newcommand{\grootfox}[1]{\textbf{\textsf{P5}}}
\newcommand{\seaonion}[1]{\textbf{\textsf{P6}}}
\newcommand{\sticklog}[1]{\textbf{\textsf{P7}}}
\newcommand{\donutweb}[1]{\textbf{\textsf{P8}}}
\newcommand{\fogalive}[1]{\textbf{\textsf{P9}}}
\newcommand{\starship}[1]{\textbf{\textsf{P10}}}
\newcommand{\toystory}[1]{\textbf{\textsf{P11}}}
\newcommand{\goldfish}[1]{\textbf{\textsf{P12}}}
\newcommand{\pumpkins}[1]{\textbf{\textsf{P13}}}
\newcommand{\sunsalad}[1]{\textbf{\textsf{P14}}}
\newcommand{\webzebra}[1]{\textbf{\textsf{P15}}}

\newcommand{\dquote}[1]{\enquote{#1}}
\newcommand{\squote}[1]{\enquote*{#1}}

\newif\ifhidefigs
\hidefigsfalse    

\ifhidefigs
  \usepackage{comment}
  \excludecomment{figure} 
\else
\fi

\newif\ifrevision
\ifrevision
  \newcommand{\rrev}[3]{\textcolor{blue}{[RevID #1: #3]}} 
\else
  \newcommand{\rrev}[3]{}
  \newcommand{\rrrev}[2]{}
\fi

\AtBeginDocument{%
  }

\setcopyright{acmlicensed}
\copyrightyear{2026}
\acmYear{2026}

\acmConference[CHI '26]{Proceedings of the 2026 CHI Conference on Human Factors in Computing Systems}{April 13--17, 2026}{Barcelona, Spain}

\begin{document}
\begin{CJK*}{UTF8}{gbsn}
\renewcommand\footnotetextcopyrightpermission[1]{} 
\settopmatter{printacmref=false} 
\title[Exploring the Potential of AI Agents for High Subjectivity Decisions in Advance Care Planning]{Words to Describe What I’m Feeling: Exploring the Potential of AI Agents for High Subjectivity Decisions in Advance Care Planning}


\author{Kellie Yu Hui Sim}
\email{kellie_sim@mymail.sutd.edu.sg}
\orcid{0009-0005-6451-7089}
\affiliation{
  \institution{Singapore University of Technology and Design}
  \city{Singapore}
  \country{Singapore}
}
\authornote{Corresponding author. This is a preprint of the paper accepted at CHI 2026. The final version will be available in the ACM Digital Library.}

\author{Pin Sym Foong}
\email{pinsym@nus.edu.sg}
\orcid{0000-0002-4437-8326}
\affiliation{
  \institution{Telehealth Core \\ National University of Singapore}
  \city{Singapore}
  \country{Singapore}
}

\author{Chenyu Zhao}
\email{cinderella.zchenyu0912@gmail.com}
\orcid{0009-0009-2502-220X}
\affiliation{
  \institution{Singapore University of Technology and Design}
  \city{Singapore}
  \country{Singapore}
}

\author{Melanie Yi Ning Quek}
\email{melanieqyn@gmail.com}
\orcid{0009-0002-4739-6834}
\affiliation{
  \institution{National University of Singapore}
  \city{Singapore}
  \country{Singapore}
}

\author{Swarangi Subodh Mehta}
\email{swarangi_mehta@sutd.edu.sg}
\orcid{0009-0008-7828-8530}
\affiliation{
  \institution{Singapore University of Technology and Design}
  \city{Singapore}
  \country{Singapore}
}

\author{Kenny Tsu Wei Choo}
\email{kenny_choo@sutd.edu.sg}
\orcid{0000-0003-3845-9143}
\affiliation{
  \institution{Singapore University of Technology and Design}
  \city{Singapore}
  \country{Singapore}
}

\renewcommand{\shortauthors}{Sim et al.}

\begin{abstract}
\input{sections/0_abstract}
\end{abstract}


\begin{teaserfigure}
  \includegraphics[width=\textwidth]{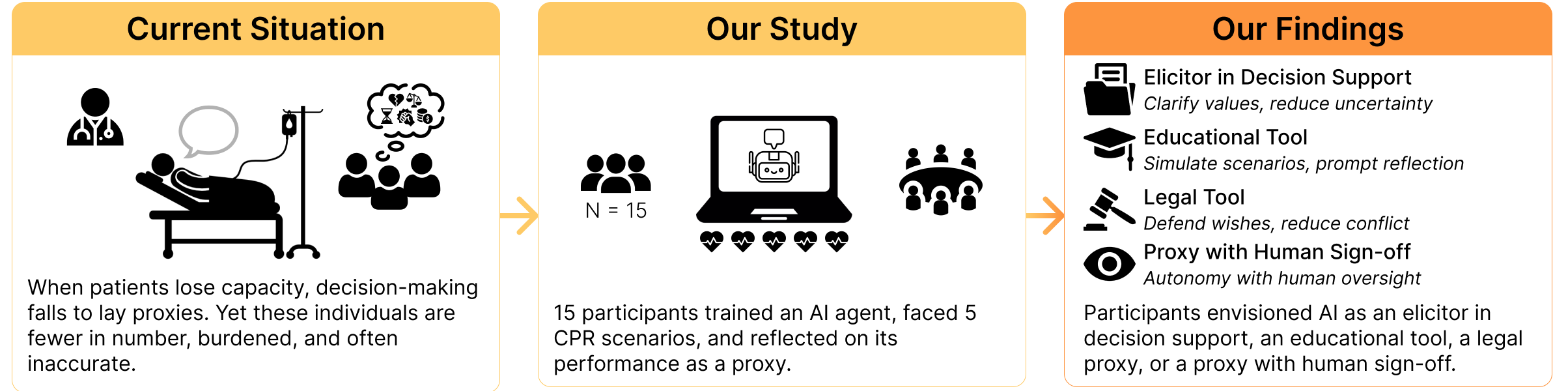}
  \caption{Summary of current challenges, study design and findings.}
  \Description{Three-panel diagram. 
  Left panel: "Current Situation" shows that when patients lose capacity, ACP relies on proxies, but caregivers are fewer, burdened, and often inaccurate. 
  Middle panel: "Our Study" shows 15 participants training an AI agent, facing five CPR scenarios, and reflecting on its proxy role. 
  Right panel: "Our Findings" shows three roles participants envisioned for AI: decision-support tool (elicits values, clarifies trade-offs), advocate (replays or defends wishes), and proxy (makes decisions on behalf).}
  \label{fig:teaser}
\end{teaserfigure}


\maketitle


\input{sections/1_introduction}
\input{sections/2_related-work}
\input{sections/3_methodology}
\input{sections/4_results}
\input{sections/5_discussion}
\input{sections/6_limitations}
\input{sections/7_conclusion}
\begin{acks}
    \input{sections/8_acknowledgements}
\end{acks}

\bibliographystyle{ACM-Reference-Format} 
\bibliography{main}

\onecolumn
\appendix
\input{appendices/scenarios}
\input{appendices/acpagent-prompts}
\input{appendices/cpr-info-sheet}
\input{appendices/participant-reflection-questions}
\input{appendices/participant-screener-form}
\input{appendices/pref-changes}


\end{CJK*}
\end{document}

%% file: sections/0_abstract.tex
Loss of decisional capacity, coupled with the increasing absence of reliable human proxies, raises urgent questions about how individuals' values can be represented in Advance Care Planning (ACP). To probe this fraught design space of high-risk, high-subjectivity decision support, we built an experience prototype (\acpagent{}) and asked 15 participants in 4 workshops to train it to be their personal ACP proxy. We analysed their coping strategies and feature requests and mapped the results onto axes of agent autonomy and human control. Our findings show a surprising 86.7\% agreement with \acpagent{}, arguing for a potential new role of AI in ACP where agents act as personal advocates for individuals, building mutual intelligibility over time. We propose that the key areas of future risk that must be addressed are the moderation of users' expectations and designing accountability and oversight over agent deployment and cutoffs.


%% file: sections/1_introduction.tex
\section{Introduction}

Societies worldwide are rapidly ageing. Longer lifespans and declining fertility have made older adults a fast-growing segment of the population, placing new demands on healthcare, caregiving, and decision-making systems~\cite{roserGlobalPopulationPyramid2019, schwarzInvertingPyramidPension2014, popeLegalFundamentalsSurrogate2012}. A central challenge arises when older adults lose decisional capacity: who should speak for them? Traditionally, family caregivers have assumed this proxy role. Yet demographic shifts have shrunk the caregiver pool, increasing the likelihood that older adults have no proxy or must rely on unfamiliar strangers, such as public guardians, for critical medical decisions~\cite{wendlerSystematicReviewEffect2011, vigSurvivingSurrogateDecisionMaking2007, shalowitzAccuracySurrogateDecision2006}.

One widely promoted way to give patients a voice before they lose capacity \cite{malhotraWhatEvidenceEfficacy2022, ooAssessingConcordancePatient2019} has been Advance Care Planning (ACP), in which individuals articulate their preferences for future end-of-life scenarios~\cite{agarwalAdvanceCarePlanning2018}. Yet completing an ACP is far from straightforward. The decisions involved--such as whether to undergo resuscitation--are high-risk (involving life and death) and highly subjective (rooted in personal values and priorities)~\cite{epsteinDevelopmentAdvanceCare2017}.

\rrev{1}{kel}{revised definition and conceptual framing of ACP in response to concerns about accuracy, flexibility, and the need to present ACP as an ongoing process (1AC/2AC)}
Although ACP is often described in strategy and policy research as an ongoing, conversation-centred process~\cite{hickmanCarePlanningUmbrella2023}, in practice it often becomes a once-off, documentation-focused event, with limited revisiting or revision over time~\cite{malhotraComplexityImplementingNationwide2022}.
Several process issues contribute: ACP documents are hard to retrieve and share across care settings, making iterative updating impractical.
For example, clinicians frequently cannot access forms completed elsewhere, creating barriers to ongoing review~\cite{lamasAdvanceCarePlanning2018}.
Public awareness is also limited--many do not know ACP documents can be updated~\cite{ngAwarenessAttitudesCommunityDwelling2017a}, and often revise them only during acute threats such as COVID-19~\cite{malaniNationalPollHealthy2021}.
Together, these factors render ACP documents into static commitments made for unpredictable futures, raising questions about their relevance for evolving clinical circumstances, health conditions, and personal priorities.

Advances in artificial intelligence (AI) have prompted interest in their role in complex, high-risk judgements. AI systems now excel at handling large datasets and generating predictions, with applications in domains such as autonomous driving and clinical diagnostics. Yet the design of AI-enhanced decision-support tools for value-laden, high-subjectivity domains like ACP remains relatively under-explored. Why and how should such systems act? Should they offer decision support, act as autonomous proxies, or adopt intermediary roles? For HCI, tackling these questions requires understanding how users conceptualise autonomy and control in AI-enhanced systems designed for this purpose.

To explore these questions, we developed \acpagent{}, an experience prototype that simulates training an AI proxy to make resuscitation decisions on users' behalf. In four workshops with 15 participants, users first interacted with the system individually and then collectively reflected on their experiences. We found a high level of alignment: participants and \acpagent{} agreed on over 80\% of resuscitation decisions. More revealing were participants’ various envisionments of \acpagent{}’s role, which ranged from a low-autonomy decision-support tool, to a high-autonomy proxy, with a third, intermediary framing of a personal advocate that learns preferences and speaks on their behalf, in care planning conversations.

We mapped these findings onto Shneiderman's~\cite{shneidermanHumanCenteredAI2022} autonomy and control framing for human-centred AI. We conceptualised participants' expressed requirements for decision support in high-risk, high-subjectivity decisions as \textit{forces} that shape desired agent roles in particular directions. From this analysis, we advanced design considerations for the most promising framing: \textbf{AI-enhanced patient advocates that combine adaptability with respect for personal values over time.}
In doing so, this paper contributes:
\begin{enumerate}
    \item \acpagent{}, a prototype system for exploring ACP decisions through training an AI agent.
    \item Empirical findings on how people perceive and interact with a trainable AI proxy in a high-risk, high-subjectivity context.
	\item \rrev{3}{kel}{revised contribution framing and strengthened articulation of limitations in current ACP processes in response to 2AC. aligned introduction and discussion so the design contribution is clearly motivated and reflects the scope of Section 5.3 (which has also been refined).} Design considerations for a new role of AI as an advocate in future care planning, towards addressing current limitations in ACP practice.
    \end{enumerate}
By situating AI within the terrain of computationally-supported ACP decision-making, our work extends HCI research on designing AI for high-risk, value-driven domains and highlights opportunities to support ageing societies where human proxies are increasingly scarce.

%% file: sections/2_related-work.tex
\section{Related Work}

\subsection{The Evolution of Delegation Systems}
\subsubsection{From Legal Documentation to Education and Conversation}
\rrev{7}{kel}{added more information on country-specific ACP and AD regulations to contextualise participant interpretations and clarify legal obligations around CPR, addressing 2AC's concerns about jurisdictional differences.}
Ageing nations like the United States, South Korea, and Finland~\cite{roserGlobalPopulationPyramid2019, schwarzInvertingPyramidPension2014, pirhonenTheseDevicesHave2020} face rising longevity and declining birthrates, often accompanied by a gradual loss of functional capacity, and increasing physical and decisional support needs~\cite{foong_prevalence_2025}.
Instruments such as advance medical directives, ACPs and Lasting Power of Attorneys (LPAs) aim to ensure patients' preferences are respected after they lose capacity, with best practices emphasising explicit conversations among patients, proxies and healthcare professionals~\cite{hickmanCarePlanningUmbrella2023, sudoreDefiningAdvanceCare2017}. 

In practice, incorporating patients' voices remains difficult due to documentation challenges~\cite{malhotraComplexityImplementingNationwide2022, kuusistoAdvanceCarePlanning2020}, problems retrieving and reviewing directives and applying their elicited preferences~\cite{morrisonWhatsWrongAdvance2021}.
Consequently, researchers increasingly argue that ACP should foreground conversation between proxies, patients, and clinicians~\cite{aldousMeasuringEngagementAdvance2020}, thereby fostering deeper understanding of patient preferences and values, and resolving conflicts with proxies' values~\cite{howardMeasuringEngagementAdvance2016, foongDesigningCaregiverfacingValues2024}.
These efforts may or may not produce documentation.

This evolution in how ACP is conceptualised and operationalised is likely one of the reasons for cross-national differences in delegation systems. 
In our study context, ACPs are not legally binding.
Also, a patient may appoint a legally binding proxy using an LPA, whose powers activate only after loss of capacity.
Unlike jurisdictions where these documents carry statutory enforcement of the patient's wishes--doctors in our country are \textbf{not} legally required to follow ACPs or LPA-appointed proxies' instructions when making resuscitation decisions, but are asked to consider patients' best interests in all decisions~\cite{menon_mental_2013}.

\subsubsection{The Limitations of Proxy Decision-Making}
Serious illness involves many unknowns. \citet{epsteinDevelopmentAdvanceCare2017} describes this as 'epistemic uncertainty', which creates proxy challenges as they balance risks across outcomes and contend with unstable prognoses. ACP decisions are hence often characterised as 'preference-sensitive': they hinge not on objective metrics, but on patients' preferences or proxies' interpretations.
So, proxies face emotional, ethical, and interpersonal challenges--broaching death, managing guilt, navigating family conflict, and judging whether past wishes apply to the present situation~\cite{foongDesigningCaregiverfacingValues2024, fleurenUnderlyingGoalsAdvance2020, suFamilyMattersEffects2014, wendlerSystematicReviewEffect2011}. Unsurprisingly, proxies misrepresent patients' wishes in about one-third of cases~\cite{shalowitzAccuracySurrogateDecision2006}. 


Delegation processes like ACP are further strained by shrinking caregiver availability in ageing nations. Without a trusted family member, responsibility may fall to healthcare providers or institutions that lack the personal knowledge of the patient, \textit{increasing} the risk that preferences are overlooked~\cite{spitzerHumanDelegationBehavior2025, suExperiencesPerspectivesFamily2020, fineEarlyExperienceDigital2016}.
As demographic shifts continue, delegation processes may not be able to only rely on caregivers to retain deep knowledge of patients' preferences.

\subsubsection{Delegation to AI}
To address these challenges, recent work has explored AI support in ACP.
~\citet{ariozArtificialIntelligencebasedApproaches2025} summarises applications into three domains: (1) extracting preferences from records; (2) predicting ACP eligibility/timing; and (3) guiding users through self-reflective conversations to articulate the patient's values and consequent care preferences.
In the last category, studies such as \textit{PreCare}~\cite{hsuPreCareDesigningAI2025} and \citet{fooBenefitsRisksLLMs2025} show how LLMs can scaffold value exploration and highlight conflicts, enhancing engagement and clarity. 

However, these tools remain advisory: they support reflection and conversation but were not designed to act as proxies in the absence of human proxies. 
We now turn to recent research on the concept of delegating decisions to AI, outlining current knowledge with respect to the scope of this paper: the delegation of authority for high-risk, high-subjectivity decisions.

\subsection{Conditions for Delegation to AI}
Research on delegating human decision-making to AI positions systems along a continuum between human control and AI autonomy~\cite{shneidermanHumanCenteredAI2022}. 
Prior work examines factors that move people along this continuum.
\citet{candrianRiseMachinesDelegating2022} suggests that when facing potential losses, humans perceive AI as lacking self-interested motives and therefore less risky and more controllable, increasing willingness to delegate--especially for low-subjectivity decisions with clear external standards of accuracy.

Other research identifies conditions that further strengthen delegation preferences. \citet{hemmerHumanAICollaborationEffect2023} showed that satisfaction and confidence increase when people understand why AI assumes decision-making responsibility in collaborative settings, while ~\citet{spitzerHumanDelegationBehavior2025} found that explicit contextual information about task demands and system competence boosts willingness to delegate.
Interestingly, in more ‘sensitive’ contexts involving potential loss or interpersonal tension, AI may be preferred as it can act as an “emotional buffer”, avoiding the social risks associated with human-led delegation~\cite{candrianRiseMachinesDelegating2022}.

\subsubsection{Nature of the Task}
The seemingly contradictory preference for AI in higher-risk situations may be explained by ~\citet{laiScienceHumanAIDecision2023}'s \emph{structure of decisions} framework, which proposes that delegation design depends on the nature of the decision itself. 
They characterise decision tasks along three dimensions: (1) task \textbf{risk} (low to high), (2) the amount of required human domain \textbf{expertise} and whether it can be solved with existing human intuitions (intuitive versus expert knowledge needed), and (3) decision \textbf{subjectivity} (extent to which personal values versus objective criteria dominate).

We define \textbf{high-risk decisions} as those involving significant financial and/or emotional prospective losses, with high reversal costs~\cite{kunreutherHighStakesDecision2002}, including situations where complex and ambiguous factors may lead decision-makers toward erroneous choices with potentially catastrophic consequences~\cite{sahohRoleExplainableArtificial2023}.
We define \textbf{high-subjectivity decisions} as those requiring contextual interpretation, where multiple possible outcomes may be equally valid~\cite{aoyaguiMatterPerspectivesContrasting2025}.


Applying \citet{laiScienceHumanAIDecision2023}'s framework, we find that prior delegation research has mainly examined lower-risk or lower-subjectivity tasks (e.g., consumer choice, hiring, financial risk prediction~\cite{hemmerHumanAICollaborationEffect2023}). By contrast, end-of-life decision-making and ACP are high-risk, involve variable human expertise, and depend on deeply subjective values. Hence, insights from low-risk, low-subjectivity research may not transfer. For example, AI's role as an "emotional buffer" arises in high-risk settings and is absent in lower-risk settings. Moreover, supporting high-subjectivity decisions remains challenging because outcome 'correctness' is best determined by subjective measure, and not by objective data alone. To our knowledge, little empirical research has addressed the use of AI in such higher-risk decisions.

\subsection{The Potential of Generative AI as Agents in High-Subjectivity Tasks}
We now turn to using generative AI as agents capable of handling subjective content such as values. Large Language Models (LLMs) 
enable conversational systems that generate context-sensitive, personalised outputs across diverse domains such as mental health~\cite{songTypingCureExperiences2025, jinDontKnowWhy2025, simUnderstandingEmotionsEngaged2024}, values elicitation~\cite{fooBenefitsRisksLLMs2025, xuGoalsActionsDesigning2025}, and reflection~\cite{songExploreSelfFosteringUserdriven2025, arakawaCoachingCopilotBlended2024, kimMindfulDiaryHarnessingLarge2024, nepalMindScapeStudyIntegrating2024}.
GenAI extends these systems by maintaining memory, modelling intentions, and acting persistently over time, opening possibilities--and risks--of machines representing people’s values~\cite{morrisGenerativeGhostsAnticipating2025} through agentic, personal advocacy-like roles.

Recent research highlights three interrelated concerns in designing such agents: (1) \textit{interactional style} (tone, persona, expressivity), which shapes trust and willingness to delegate~\cite{okosoExpressionsChangeDecisions2025, leyerMEYOUAI2019}; (2) \textit{delegation and authority}, particularly how framing and transparency of the system affects user confidence~\cite{spitzerHumanDelegationBehavior2025, candrianRiseMachinesDelegating2022}; and (3) \textit{representation and identity}, including how agents preserve or simulate aspects of the self across time~\cite{morrisGenerativeGhostsAnticipating2025, khotChallengingFuturesUsing2025}.

Non-experts are particularly swayed by affective framing and communicative style, with "future self" personas enhancing emotional resonance~\cite{okosoExpressionsChangeDecisions2025, khotChallengingFuturesUsing2025}. Across expertise levels, willingness to rely on agents increases when contextual information and system competence are clearly conveyed~\cite{spitzerHumanDelegationBehavior2025, candrianRiseMachinesDelegating2022, leyerMEYOUAI2019}.

Legal and ethical debates about the limits of proxy authority add nuance to these dynamics.
Current proxy decision-making policy emphasises fidelity to patient wishes~\cite{popeLegalFundamentalsSurrogate2012}, yet human proxies often reinterpret or override preferences under stress or conflict~\cite{bandiniIfWeTurned2021}.
This raises questions on the limits of agentic proxy authority, required auditing, and how AI agents should balance autonomy with contextual judgment--acting conservatively, assertively, or neutrally.

In summary, while interest in AI systems that represent or preserve human values is growing, research on how such systems could operate as proxies remains nascent. 
This study, therefore, examines the design of an AI agent to act as a proxy for the patient in ACP, specifically aiming to: 

\begin{enumerate}
    \item Explore the users' experience of using an AI agent to make high-risk, high-subjectivity decisions.
    \item Examine this experience for potential roles and framings of an AI agent in supporting such decisions. 
\end{enumerate}

%% file: sections/3_methodology.tex
\section{Methodology}
To examine how people engage with an AI-powered agent designed to serve as a patient proxy in ACP, we developed an LLM-powered system (\acpagent{}) that simulates training such an agent to be one's own proxy in ACP decisions. 
We then conducted a series of participatory workshops with individuals who may one day rely on or interact with such a system.
In these workshops, participants trained \acpagent{} and reflected on their experience 
The study received Institutional Review Board approval (SUTD IRB-25-00727), and all procedures followed ethical research guidelines.

\begin{table*}[htbp!]
\centering
\small
\caption{Participant characteristics across the four workshops (WS). 
The table summarises demographic factors (age, gender), relevant prior experiences (delegating authority, e.g., ACP/LPA, being able to identify a trusted other, personal experience with disease), and digital orientation (use of digital health tools, AI, IT familiarity). 
These factors shaped how participants engaged with \acpagent{} and informed their reflections on its role in decision-making.}
\label{tab:participants}
\renewcommand{\arraystretch}{1.12}
\begin{tabular}{P{0.3cm} P{0.3cm} P{0.3cm} P{1.1cm} P{2.5cm} P{1.8cm} P{1.8cm} P{1.7cm} P{1.3cm} P{2.1cm}}
\toprule
\textbf{WS} & \textbf{ID} & \textbf{Age} & \textbf{Gender} & \textbf{Experience with Authority Delegation (60\% No)} & \textbf{Can Identify Trusted Other (40\% No)} & \textbf{Personal Experience with Disease} & \textbf{Use of Digital Health Tools (73.3\% No)} & \textbf{Use of AI (26.7\% No)} & \textbf{IT Familiarity Score (Out of 60) } \\
\midrule
1 & \baritone{}  & 45 & Female & No  & No  & Other & No  & Regularly    & 29 \\
1 & \redpasta{}  & 58 & Female & No  & Yes & Self  & Yes & Occasionally & 35 \\
1 & \trombone{}  & 48 & Male & No  & No  & Both  & No  & Occasionally & 28 \\
\hdashline
2 & \cowalive{}  & 41 & Female & Yes & No  & Other & Yes & Occasionally & 26 \\
2 & \grootfox{}  & 67 & Male  & Yes & Yes & Self  & No  & Occasionally & 29 \\
2 & \seaonion{}  & 60 & Female & Yes & Yes & Both  & No  & No               & 33 \\
2 & \sticklog{}  & 51 & Male  & Yes & Yes & Other & Yes & Occasionally & 32 \\
\hdashline
3 & \donutweb{}  & 49 & Male   & No  & No  & Other & No  & Regularly    & 36 \\
3 & \fogalive{}  & 51 & Female & Yes & Yes & Other & No  & No               & 46 \\
3 & \starship{}  & 51 & Male   & No  & Yes & Both  & No  & No               & 41 \\
3 & \toystory{}  & 47 & Female & No  & Yes & Other & No  & No               & 35 \\
\hdashline
4 & \goldfish{}  & 62 & Female & No  & No  & Other & Yes & Occasionally & 27 \\
4 & \pumpkins{}  & 68 & Male   & Yes & No  & Other & No  & Occasionally & 33 \\
4 & \sunsalad{}  & 54 & Male   & No  & Yes & Other & No  & Regularly    & 26 \\
4 & \webzebra{}  & 45 & Male   & No  & Yes & Both  & No  & Occasionally & 34 \\
\bottomrule
\end{tabular}
\par\vspace{3pt}\footnotesize
\textit{Notes:} 
"IT Familiarity Score" = sum of 20 technology familiarity items (lower = more frequent use; 1 = Daily use, 2 = Seldom use, 3 = Never use).  
\end{table*}

\FloatBarrier
\subsection{Participants}
\rrev{13}{kel}{adjusted clarity of participant demographics reporting (mean, SD for age)}
We recruited 15 participants (7 female, all aged 40 and above, $mean = 53.1$, $SD = 7.9$; see Table~\ref{tab:participants}) via social media posters targeting people with personal or family experience of life-limiting illness (e.g., advanced cancer, stroke, dementia, heart disease).
Each participant received $\sim$USD23 compensation.
The group was well-educated, with most holding a Bachelor's degree, and most employed full- or part-time ($n=11$).
Awareness of delegation systems was high: 13 participants (86.7\%) had prior exposure to ACP, typically through hearing about it or acting as donor or proxy, and all were aware of LPA.
6 (40\%) had direct experience as donors or donees in ACP/LPA contexts, and an equal number could not identify a trusted proxy (i.e., unable to name someone for LPA, responsible for acting per LPA, or another trusted appointee, such as a friend).
Most reported personal experience with life-limiting illness(es), with 6 experiencing it personally, and 13 (86.7\%) having close family or friends affected.

\rrev{2}{kel}{clarified "digital health tools" definition, as per R2's comment}
We assessed digital literacy by examining prior use of digital health tools for health-related decision-making.
Use varied: 11 participants had used digital or AI tools for health decision-making, whereas 4 had not.
A 20-item digital behaviour inventory adapted from Geyer~\cite{faettTeachingSelfManagementSkills2013} reflected familiarity with routine digital tasks such as online search, with AI chatbots primarily for information seeking (60\% daily, 40\% occasional).

\subsubsection{Participant Safety}
Given the sensitivity of ACP discussions, we exercised care in recruitment and facilitation.
Recruitment materials noted that the study involved decision-making about life-limiting illness and end-of-life planning, ensuring informed participation.

At the start of each session, we reiterated voluntary participation, the right to pause, skip, or withdraw at any time, and that disclosing personal health conditions was optional. 
We monitored for discomfort, offered breaks and refreshments, and conducted structured debriefs. 
An information sheet with psychosocial support hotlines was also provided.
While no participants chose to stop or withdraw, safety measures were consistently emphasised.

\subsection{Implementation}
All interactions were delivered via a custom-built web application developed in React\footnote{https://react.dev/} hosted on Amazon Web Services\footnote{https://aws.amazon.com/}, with the underlying LLM being GPT-4o, accessed via the OpenAI API\footnote{https://openai.com/api/}.
Each round consisted of two form-like screens: (1) preference setup; and (2) scenario-based decision-making and interaction with \acpagent{}. 

\subsubsection{Training Scenarios}\label{sec:training_scenarios}

Participants interacted with \acpagent{} across five CPR-related scenarios (see Appendix~\ref{appendix:scenarios} for full scenarios):

\begin{itemize}
    \item \textbf{Scenario 1 - Recoverable Episode:} A sudden but treatable health event (e.g., infection) where CPR would be performed by default if the heart stopped.
    \item \textbf{Scenario 2 - Terminal Illness Becomes Clear:} Cancer spread made the illness incurable, though still relatively well, with an estimated 1-2 years of life.
    \item \textbf{Scenario 3 - Severe Side Effects from Aggressive Treatment:} Bedbound after chemotherapy with limited quality of life and required assistance with daily activities.
    \item \textbf{Scenario 4 - Cost and Resource Burden:} Prolonged dependency leading to nursing home placement, raising concerns about financial and care burdens.
    \item \textbf{Scenario 5 - Limited Life Expectancy with Family Conflict:} Weeks left to live and reduced consciousness, relatives disagreeing on whether to pursue CPR, and unclear prior wishes.
\end{itemize}

Scenarios were designed to cover a wide range of diseases and prognostic severity.
The first two depicted better prognoses, 
while the latter three involved advanced decline.
Each required them to indicate whether they wanted CPR under changing health trajectories and circumstances.
While other ACP decisions (e.g. tube feeding) exist, we focused on CPR because it is widely recognised, emotionally salient, and frequently included in ACP documentation as a complex yet concrete life-sustaining treatment~\cite{winterPatientValuesPreferences2013}.

In designing this study, we anticipated two risks: that the scenarios might not generate sufficient reflection or decisional dilemma, and that participants might overlook the scenario text entirely. To verify that the scenarios functioned as intended, we included a self-report check asking participants whether the scenario had influenced their decision.

\subsubsection{Simulating Agent Training}
\acpagent{} was designed to simulate a trainable ACP proxy, with participants in the role of "trainer" and \acpagent{} as "trainee".
We chose to simulate the experience of training by feeding participants' inputs (from each round, as a JSON file) into subsequent prompts to act as a technology probe~\cite{hutchinsonTechnologyProbesInspiring2003}, especially since model training or fine-tuning would not result in real-time interactions.
For example, if a participant rated their tolerance for cost as 'low' or disagreed with a recommendation, this information was included in the next prompt to shape \acpagent{}'s reasoning.

To reduce risks of over-trust and information overload in LLM systems~\cite{fooBenefitsRisksLLMs2025}, participants made their own CPR choices \textit{before} seeing \acpagent{}'s suggestion.
To support transparency in AI reasoning~\cite{haoAdvancingPatientCenteredShared2024}, \acpagent{} displayed its reasoning in context, offering a lower-risk space for reflecting on personal priorities~\cite{fooBenefitsRisksLLMs2025}.
Prompts instructed the model to return four labelled sections: (1) Thinking Process, (2) Recommended Decision, (3) Reasons, and (4) Binary Decision (Yes/No CPR) (see Appendix~\ref{appendix:acpagent-prompts} for full prompt templates and sample outputs).

\subsection{Workshop Protocol}
At the start of the workshop, participants were briefed as follows: \dquote{This study aims to understand how individuals like you would make important healthcare decisions if you no longer had the capacity to do so, or if you do not have any caregiver support.}
They were also told: \dquote{We are interested in participants' views on AI-powered tools, including their comfort levels, expectations, and concerns about using such tools to advocate for their needs or represent their values.}
Participants were asked to treat the interaction as training an AI system that could later speak on their behalf in sensitive healthcare contexts.

\begin{figure*}[htbp!]
    \centering
    \includegraphics[width=0.95\linewidth]{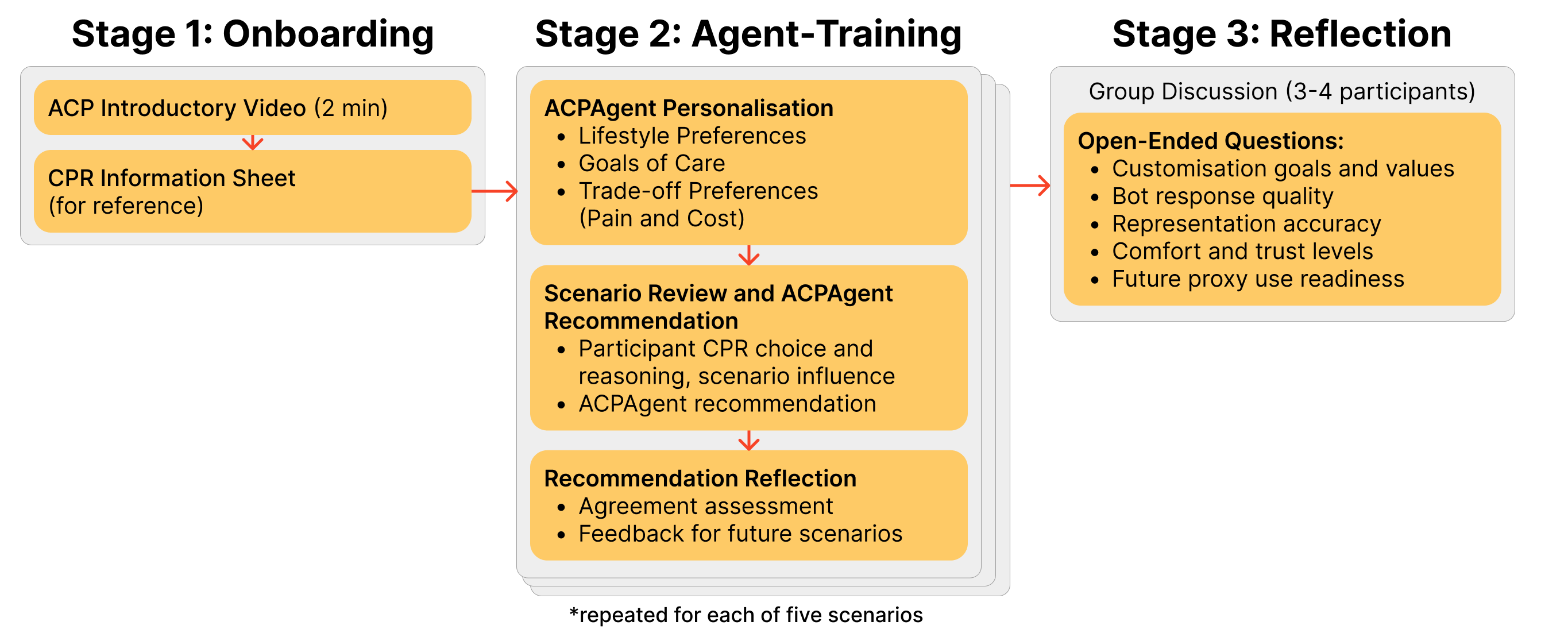}
    \caption{\textit{Three-stage workshop protocol}. 
    \textbf{Stage 1 (Onboarding):} Participants receive ACP education through a short video and CPR information sheet. 
    \textbf{Stage 2 (Agent-Training):} Across five CPR scenarios of increasing complexity, participants make initial CPR decisions, review \acpagent{} recommendations, and provide reflections on agreement, reasoning, and areas for improvement. 
    \textbf{Stage 3 (Reflection):} Groups of 3-4 participants engage in structured discussion covering customisation goals, bot response quality, representation accuracy, comfort and trust, and readiness for proxy use.}
    \Description{Flow diagram showing the three-stage research protocol. 
    Stage 1 (Onboarding): participants watch a 2-minute ACP video and review a CPR information sheet. 
    Stage 2 (Agent-Training): repeated for five scenarios, each involving scenario review, participant CPR choice and reasoning, agent recommendation, reflection, agreement assessment, and feedback. 
    Stage 3 (Reflection): groups of 3-4 participants discuss customisation goals, bot response quality, representation accuracy, comfort and trust, and readiness for proxy use.}
    \label{fig:study-overview}
\end{figure*}

Workshops comprised three stages (see Figure~\ref{fig:study-overview}).
In \textbf{Stage 1 (\textit{onboarding})}, participants watched a two-minute publicly available ACP video~\cite{aicsingaporeAdvanceCarePlanning2022} and reviewed a CPR information sheet (see Appendix~\ref{appendix:cpr-info-sheet}) to establish a shared baseline understanding.

\begin{figure}[htbp!]
  \centering
  \begin{subfigure}{\linewidth}
    \includegraphics[width=\linewidth]{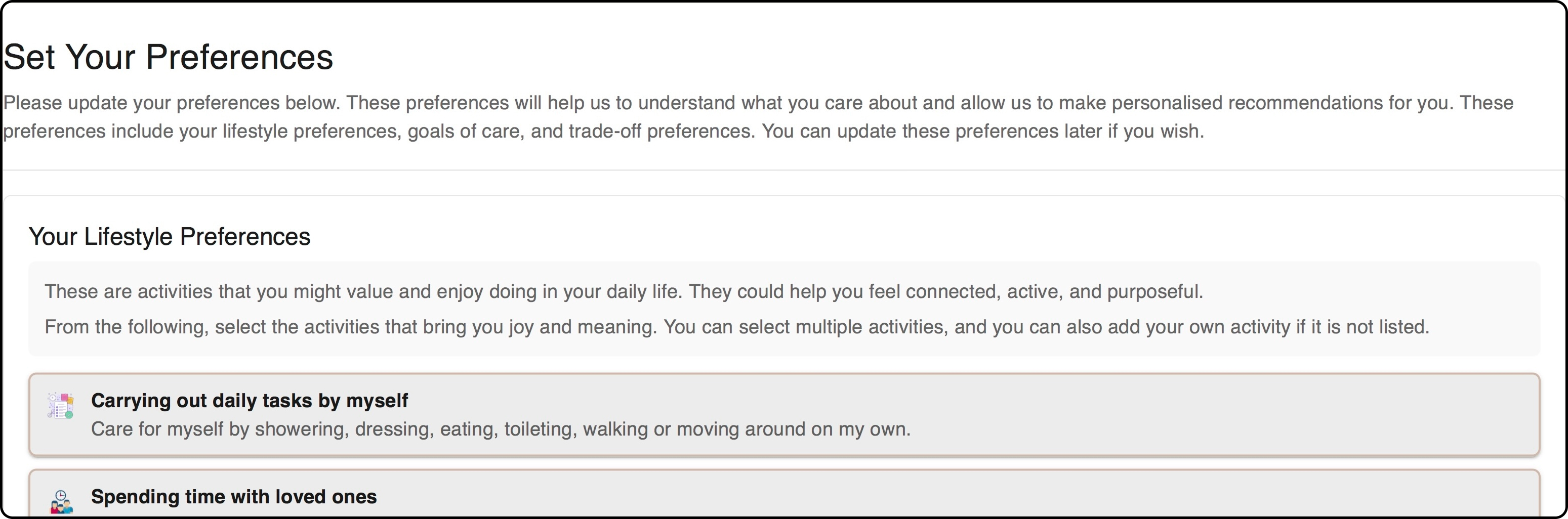}
    \caption{Lifestyle preferences}
  \end{subfigure}
  \vfill
  \begin{subfigure}{\linewidth}
    \includegraphics[width=\linewidth]{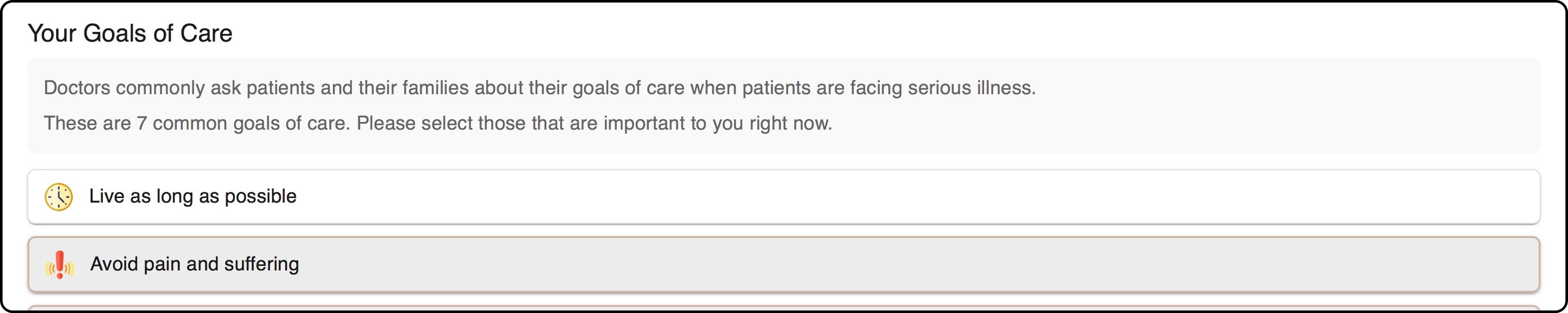}
    \caption{Goals of care}
  \end{subfigure}
  \vfill
  \begin{subfigure}{\linewidth}
    \includegraphics[width=\linewidth]{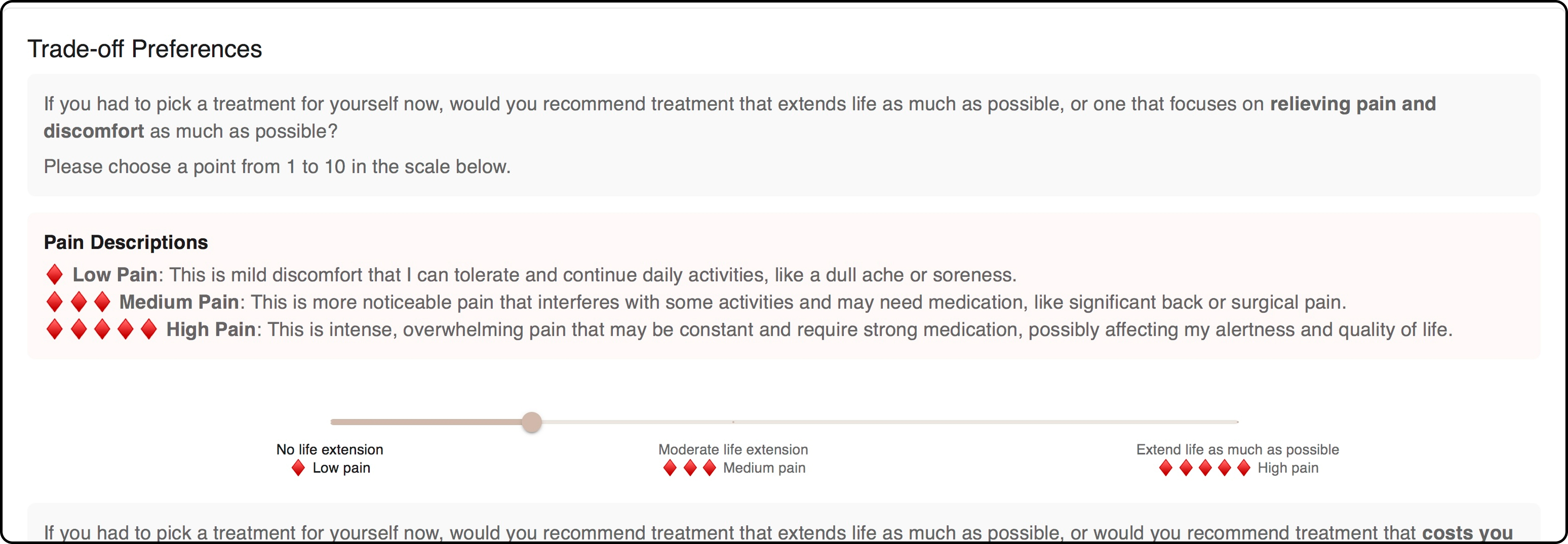}
    \caption{Trade-off preferences}
  \end{subfigure}
  \caption{Preference setup interface at the start of Stage 2 (Agent-Training). 
  Screens guided participants through configuring (a) lifestyle values, (b) goals of care, and (c) trade-off ratings (1-10) to initialise their personalised \acpagent{}.}
  \Description{Three screenshots of the preference setup interface. 
  (a) shows a list of lifestyle preferences, such as spending time with loved ones and exercising. 
  (b) shows seven common goals of care, such as living as long as possible or avoiding pain. 
  (c) shows a 1-10 slider for life extension versus pain tolerance; a similar slider was used for cost.}
  \label{fig:pref-setup}
\end{figure}

At the beginning of \textbf{Stage 2 (\textit{agent-training})}, participants were told that their inputs (e.g., values, choices, reflections) would directly inform \acpagent{}'s subsequent outputs, with the system considering only the most recent round of inputs rather than a cumulative history across rounds. 
Participants were also reminded of this across rounds where required.
Participants then specified three sets of preferences (see Figure~\ref{fig:pref-setup}):
\begin{itemize}
    \item \textbf{Lifestyle Preferences.} Participants selected meaningful daily activities taken from a national ACP resource~\footnote{https://www.aic.sg/care-services/acp-resources/} (e.g., independence, time with loved ones, exercise, work), with an option to add their own.
    \item \textbf{Goals of Care Preferences.} Participants selected from seven common ACP goals~\cite{kohNavigatingAdvancedIllness2025} (e.g., avoiding suffering, prolonging life, avoiding being a burden), reflecting frameworks used in ACP discussions.
    \item \textbf{Trade-Off Preferences.} Participants rated, on two 1-10 scales adapted from \citet{malinUnderstandingCancerPatients2006}, how they would balance life extension against (1) pain/discomfort and (2) financial cost, each anchored by descriptive examples for low/medium/high to quantify one's trade-off between quality and cost of life, relative to a longer lifespan.
\end{itemize}

\begin{figure}[htbp!]
    \centering
    \begin{subfigure}{\linewidth}
        \centering
        \includegraphics[width=\linewidth]{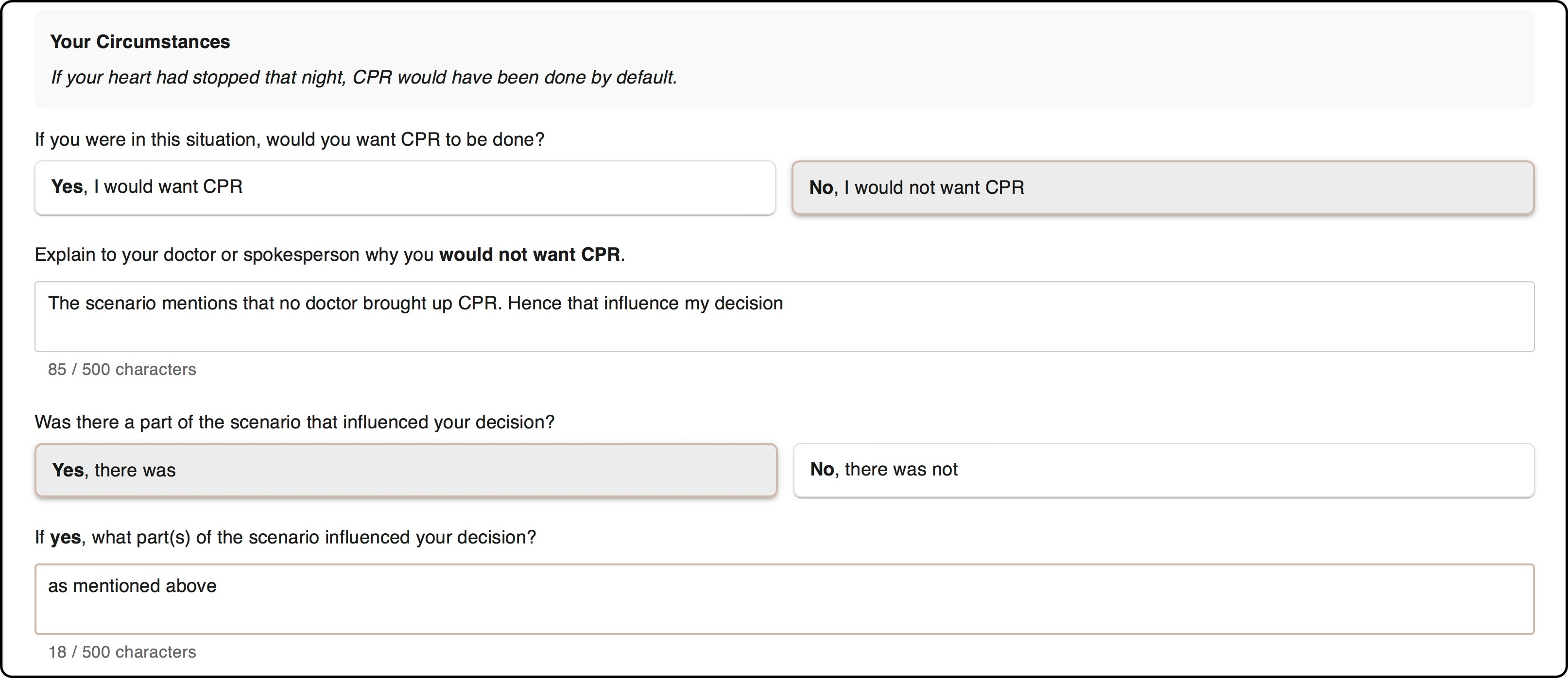}
        \caption{Scenario screen showing participant’s initial decision and whether the scenario influenced the participant's decision.}
        \label{fig:scenario-interface-a}
    \end{subfigure}
    \vfill
    \begin{subfigure}{\linewidth}
        \centering
        \includegraphics[width=\linewidth]{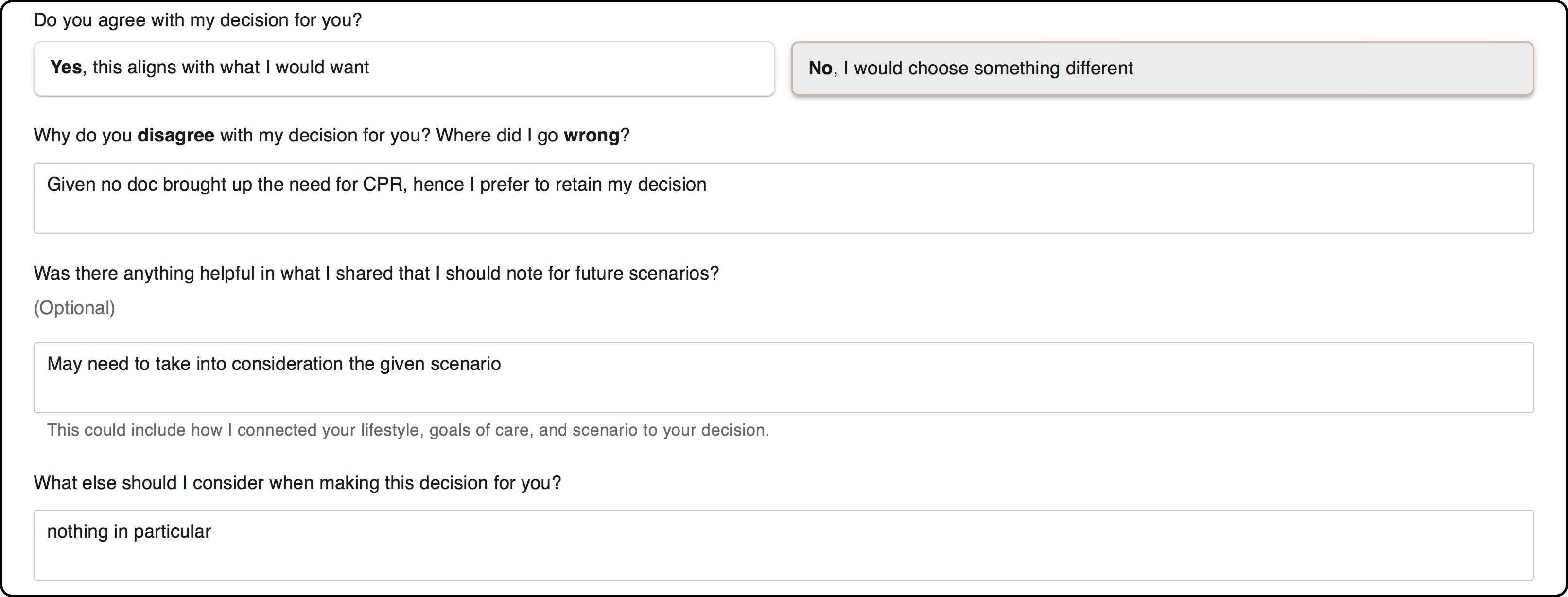}
        \caption{Screen showing reflection questions, presented after participants viewed \acpagent{}'s recommendation and reasoning breakdown.}
        \label{fig:scenario-interface-b}
    \end{subfigure}
    \caption{Scenario-based interaction interface in Stage 2 (Agent-Training). Together, these steps simulated delegated decision-making while maintaining transparency of reasoning.}
    \Description{Two screenshots of the scenario interface. 
    (a) shows a scenario description with fields for participants to indicate their CPR decision, explain their reasoning, and note whether scenario details influenced them. 
    (b) shows fields where participants can agree or disagree with the recommendation and provide feedback for future scenarios, after viewing \acpagent{}'s recommendation and structured reasoning breakdown.}
    \label{fig:scenario-interface}
\end{figure}

After setup, participants engaged in five rounds.
In each round, participants first viewed a scenario and recorded their CPR decision and reasoning (see Figure~\ref{fig:scenario-interface}).
Then, participants clicked a button to reveal \acpagent{}’s recommendation with a structured breakdown.
Participants then answered reflection questions to indicate their agreement/disagreement and provide feedback.
These responses, together with lifestyle, goals of care, and trade-off values, were used in the LLM prompt to generate \acpagent{}'s recommendations for the subsequent round.

Finally, in \textbf{Stage 3 (\textit{reflection})}, participants engaged in group discussions on alignment, comfort, and trust (see Appendix~\ref{appendix:participant-reflections} for reflection questions). 

\subsection{Data Collection}
Before the workshop, participants completed a screener on demographics, delegation and life-limiting illness experiences, presence of a trusted other, and AI and IT familiarity (see Appendix~\ref{appendix:participant-screener}).
During the workshop, three researchers observed (one for every one or two participants), took notes, and consolidated findings after each session.
Workshops were audio-recorded, transcribed using WhisperX~\cite{bain2022whisperx} and verified by two researchers.
System inputs of lifestyle preferences, goals of care, trade-off ratings for life-extensions, CPR decisions, agreements wth \acpagent{} recommendations, scenario influence counts, and free-text reflection lengths) were logged in JSON.

\FloatBarrier


\subsection{Data Analysis}
\rrev{6}{kel}{expanded the RTA methodology description in response to reviewer concerns, clarifying epistemological stance, analytic level, analyst roles, and how thematic interpretation informed the subsequent autonomy-control mapping.}
JSON data, comprising open-ended text from the CPR decision rationales, scenario elements influencing decisions, and reasons for agreement/disagreement with \acpagent{}, was analysed for cross-scenario behaviour to track preference shifts across the five scenarios.
We also analysed \textbf{Stage 3 (\textit{reflection})} responses, desired improvements to \acpagent{}, and other considerations required for decision-making.
We analysed the qualitative data using an applied thematic approach~\cite{guestAppliedThematicAnalysis2011, braunUsingThematicAnalysis2006}, combining thematic coding and interpretive mapping.
Themes were treated as analytic constructs created by the researchers to organise patterns of shared meaning relevant to the study aims, rather than as entities assumed to pre-exist within the data.

The analysis was conducted from a contextualist epistemological stance, recognising that participants' reflections were shaped by the workshop setting, scenario structure, and group dynamics.
The researchers focused on semantic meaning, staying close to participants’ expressed reasoning, while acknowledging the active role of the researchers in constructing and naming themes.

Initial open coding was conducted on individual transcripts, field notes, and reflection logs by two team members trained in qualitative methods (i.e., primary author and another researcher).
These researchers were involved in the study design and facilitation, which informed their sensitivity to the interactional and contextual aspects of the data.
Codes were iteratively reviewed, discussed, and reorganised through regular analytic meetings, during which candidate themes were actively constructed, refined, and sometimes collapsed to better capture shared meanings across participants.
We retained a larger set of analytically useful themes to reflect the diversity and nuance of participants' considerations across scenarios and discussion phases.
These themes functioned as organising lenses for reporting findings.

Subsequently, all authors reviewed the results and conducted a secondary interpretive mapping to examine how participants’ discussions aligned with \citet{shneidermanHumanCenteredAI2022}'s framework of autonomy and control.
This mapping constituted an additional layer of interpretation, explicitly linking the results to existing conceptual frameworks.

%% file: sections/4_results.tex
\section{Results}
\rrev{5}{kel}{revised and streamlined Section 4 (Results) in response to concerns (2AC, R2) about length, density, and clarity. Condensed 4.1, improved narrative flow, reorganised paragraphs, and refined or removed figures to emphasise clearer presentation of findings for both technical and ethical audiences. Added a new figure to summarise participants' activity across all rounds.}
This section presents two sets of findings. 
Firstly, we assess if the experience prototype sufficiently simulated the sense of training an agent to make proxy decisions. To do so, we present participants' behavioural patterns showing how they used \acpagent{} across scenarios.
Secondly, we present the thematic analysis of participants’ responses to \acpagent{}. 
These are presented as four envisioned roles, positioned along axes of human control and agent autonomy.


\subsection{Participants' Proxy Training Behaviour}
\rrev{10}{kel}{clarified how reflective and corrective reasoning were distinguished in the analysis, and explained how system design shaped opportunities for corrective feedback in the limitations section, addressing concerns (1AC, R2) about analytic validity.}
All 15 participants completed five decision-making rounds with \acpagent{}, each prompted by a new scenario.
To examine how actively participants engaged in training \acpagent{}, we explored three observable behaviours captured in the JSON logs:
\begin{enumerate}
    \item reporting if the scenario influenced their reasoning;
    \item changing preference settings between scenarios; 
    \item giving feedback aimed at steering \acpagent{}; 
\end{enumerate}
We treated preference edits and instructive feedback (items 2 and 3) as \textbf{corrective} uses of the system--explicit attempts to adjust how \acpagent{} would reason in subsequent rounds. 
By contrast, agreement or disagreement without subsequent edits, but accompanied by written reflections on the scenario or \acpagent{}’s reasoning, was considered \textbf{reflective} use.

\begin{figure*}
    \centering
    \includegraphics[width=1\linewidth]{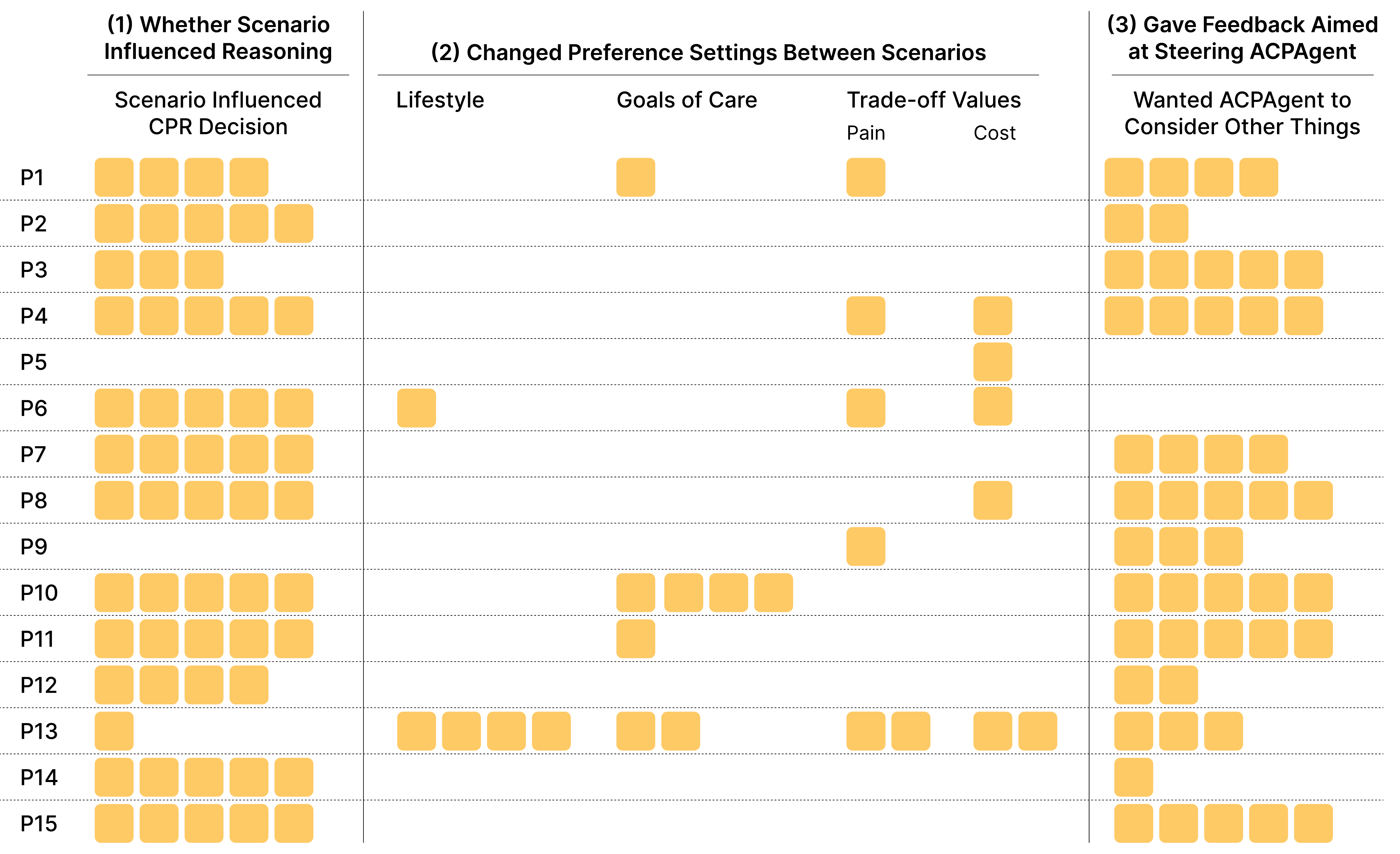}
    \caption{A summary of how frequently each of the 15 participants engaged in the 3 forms of training behaviour when interacting with \acpagent{} across the 5 scenarios: reporting whether the scenario influenced their reasoning, editing their preferences between rounds and providing steering feedback. This activity grid highlights the substantial variation in how actively participants shaped \acpagent{} across the study.}
    \Description{A participant-by-round activity grid showing when each participant engaged in specific training behaviours while interacting with \acpagent{} across the five scenarios. For each participant, the chart marks the frequency of the (1) rounds in which they reported being influenced by the scenario, (2) rounds in which they changed preference settings (lifestyle, goals of care, pain trade-off, or cost trade-off) after a scenario, and (3) rounds in which they gave feedback to steer \acpagent{}. The distribution of marked rounds illustrates the diversity of engagement patterns, with some participants being highly active across multiple categories and others making only occasional adjustments.}
    \label{fig:training-grid}
\end{figure*}

Engagement varied substantially, with some participants (e.g., \seaonion{}, \starship{}, \pumpkins{}) using all three options, frequently adjusting preferences, offering corrective instructions, and grounding their decisions in scenario details. 
Others interacted minimally (e.g., \redpasta{}, \fogalive{}, \goldfish{}), making few edits and little feedback beyond responding to \acpagent{}’s recommendations and reasoning, though each demonstrated at least one corrective behaviour (i.e., adjust inputs / provide additional context after a scenario). 
Figure~\ref{fig:training-grid} summarises how frequently each participant engaged in these behaviours.


\subsubsection{Scenario Influence}
Across the 75 decisions, participants attributed 57 (76\%) to scenario details. All but \grootfox{} and \fogalive{} reported influence at least once, but their answers in Section 4.2 suggest that they were still meaningfully engaging with the discussion about the envisioned use of \acpagent{}.
Scenarios 2, 3, and 5, involving prognosis, independence, and fluctuating consciousness, seemed particularly evocative of decisional dilemmas, producing a mix of CPR decisions (22 for, 35 against).
Scenario elements, such as patient prognosis, independence, consciousness, and financial strain, were reported to have influenced the participants' decisions. These elements were considered when participants adjusted what \acpagent{} ought to prioritise in accepting or rejecting CPR.

Even participants who rarely edited their preferences still used scenario reasoning as an indirect form of training by signalling which contextual factors they felt were salient for future rounds. 
In the participant reflection stage, participants additionally shared that it was the narrativised, sequential building of scenario complexity that helped to deepen their reflections.
As \cowalive{} later noted during the reflections phase, unlike a multiple-choice ACP form, the scenarios revealed how \dquote{minute nuances} could shift their choice between rounds.

\subsubsection{Preference Changes: Lifestyle, Goals of Care, Pain/Cost Trade-offs}
\begin{figure}[htbp!]
    \centering
    \includegraphics[width=0.75\linewidth]{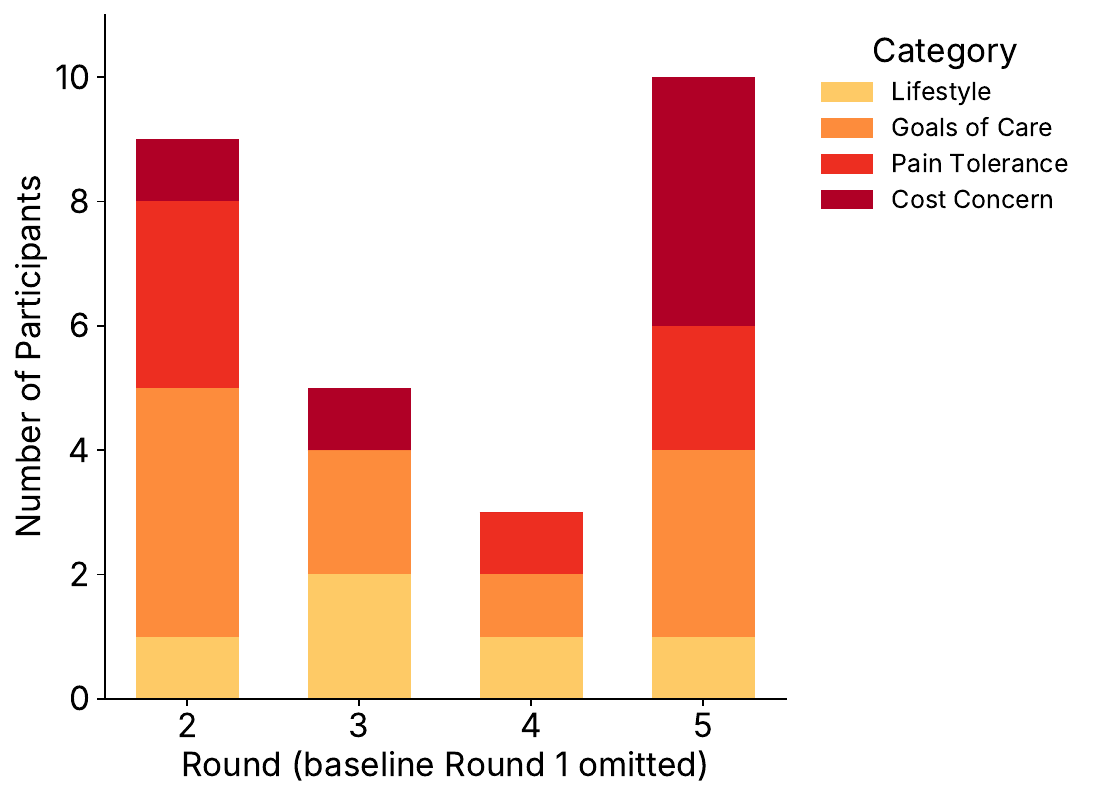}
    \caption{\textit{Preference changes by category before each round} (e.g., changes before Round 2 were made after seeing Scenario 1). \textit{Lifestyle preferences} remained stable, whereas \textit{Goals of Care} and \textit{Pain} and \textit{Cost} trade-offs showed greater variability. 
    Round 1 represents the initial baseline and is not shown.}
    \Description{Stacked bar chart showing the number of participants revising Lifestyle, Goals of Care, Pain Tolerance, or Cost Concern preferences prior to Rounds 2-5. Lifestyle remained most stable, while Goals of Care and the numeric trade-offs showed higher variability.}
    \label{fig:distribution-changes}
\end{figure}

Some participants used preference changes not simply to reflect shifting views, but as deliberate signals for steering \acpagent{}’s reasoning.
\rrev{11}{kel}{added short explanation on why new dilemmas emerged when scenarios 1 and 4 were introduced, as per R3's comment}
Lifestyle values were generally stable, while goals of care and pain/cost trade-offs showed more variability, especially after scenarios that introduced new dilemmas (e.g., reversibility, dependency, or affordability in Scenarios 1 and 4) in their descriptions (see Appendix~\ref{appendix:pref-changes}).
9 of 15 participants (60\%; \baritone{}, \textbf{\textsf{P4-6}}, \textbf{\textsf{P8-11}}, \pumpkins{}) revised at least 1 preference across the 5 rounds (Figure~\ref{fig:distribution-changes}).
For example, in the reflection questions, \cowalive{}, who lowered both pain and cost trade-offs from 5 to 1 after Scenario 4, explained that CPR would be a \dquote{burdensome life-prolonging measure}.

\subsubsection{Feedback to Steer \acpagent{}}
Towards the end of each round, participants evaluated if they agreed with \acpagent{}'s recommendation, and then were asked to provide instructions to help \acpagent{} improve.
Table~\ref{tab:agreement-table} shows a summary of these varied agreements.
In 65 of 75 cases (86.7\%), participants evaluated that they had agreed with \acpagent{}’s CPR recommendation. Among the \emph{disagreements}, (13.3\%) participants indicated that this was because the \acpagent{} had overlooked important considerations, such as family emotions, financial burden, or realistic expectations of recovery. 
For two participants (\grootfox{}, \toystory{}), disagreement led them to revise their preferences for the next round.

In response to questions about what \acpagent{} could improve on, participants inputted additional factors for \acpagent{} to consider, such as family impact, affordability, lucidity, frailty, life stage, or systemic constraints. 
These elaborations suggested that participants attempted to shape not only \acpagent{}'s outputs but its underlying decision logic. These additional feedback channels were used by participants to fine-tune what \acpagent{} should emphasise.

Among the disagreements, there were 9 cases (12\%), where \acpagent{} recommended the opposite of what the participant had chosen, but participants’ reflective feedback indicated that they now agreed with the \acpagent{}'s reasoning. 
This was an unexpected behaviour, as it suggested that \acpagent{}'s recommendations influenced participants' positions.
It seemed that the detailed reasoning could serve as a reflective mirror to participants, even when the CPR decision differed. It suggested that the influence was bi-directional, and not just human to machine. 

\begin{table*}[htbp!]
\centering
\rrev{4}{kel}{revised Table 2 and Figure 10 for accessibility and black-and-white readability as per 2AC's comment, adding clearer legends, redundant cues, and distinguishable boundary markers.}
\caption{Participant (P) vs.\ \acpagent{} (A) CPR decisions by round.
Cells are reported as dyads of (Participant / \acpagent{}).
For example, (\textcolor{agreegreen}{\ding{51}} / \textcolor{agreered}{\ding{53}}) indicates that the participant wanted CPR, but \acpagent{} recommended No CPR. 
Cell marking reflects three cases: 
(1) Unshaded: full alignment between the participant and \acpagent{};
(2) Shaded: mismatch between the participant's CPR choice and \acpagent{}'s recommendation, and the participant reported that \acpagent{}'s decision \textit{did not} align with their own;
(3) Shaded and underlined: mismatch in CPR decisions, but the participant nonetheless reported that \acpagent{}'s decision \textit{did} align with their own.
} 
\begin{tabular}{l*{5}{>{\centering\arraybackslash}p{1.5cm}}}
\toprule
\textbf{Participant (P)} & \multicolumn{5}{c}{\textbf{Round}} \\
\cmidrule(lr){2-6}
 & R1 (P/A) & R2 (P/A) & R3 (P/A) & R4 (P/A) & R5 (P/A) \\
\midrule
\baritone{} 
& \textcolor{agreegreen}{\ding{51}} / \textcolor{agreegreen}{\ding{51}} 
& \colorbox{mismatchamber}{\underline{\textcolor{agreered}{\ding{53}} / \textit{\textcolor{agreegreen}{\ding{51}}}}} 
& \colorbox{mismatchamber}{\textcolor{agreegreen}{\ding{51}} / \textcolor{agreered}{\ding{53}}} 
& \textcolor{agreegreen}{\ding{51}} / \textit{\textcolor{agreegreen}{\ding{51}}} 
& \textcolor{agreegreen}{\ding{51}} / \textit{\textcolor{agreegreen}{\ding{51}}} \\

\redpasta{} 
& \colorbox{mismatchamber}{\textcolor{agreered}{\ding{53}} / \textit{\textcolor{agreegreen}{\ding{51}}}} 
& \textcolor{agreered}{\ding{53}} / \textcolor{agreered}{\ding{53}} 
& \textcolor{agreered}{\ding{53}} / \textcolor{agreered}{\ding{53}} 
& \textcolor{agreered}{\ding{53}} / \textcolor{agreered}{\ding{53}} 
& \textcolor{agreered}{\ding{53}} / \textcolor{agreered}{\ding{53}} \\

\trombone{} 
& \textcolor{agreegreen}{\ding{51}} / \textit{\textcolor{agreegreen}{\ding{51}}} 
& \textcolor{agreegreen}{\ding{51}} / \textit{\textcolor{agreegreen}{\ding{51}}} 
& \colorbox{mismatchamber}{\textcolor{agreegreen}{\ding{51}} / \textcolor{agreered}{\ding{53}}} 
& \textcolor{agreegreen}{\ding{51}} / \textit{\textcolor{agreegreen}{\ding{51}}} 
& \colorbox{mismatchamber}{\textcolor{agreered}{\ding{53}} / \textit{\textcolor{agreegreen}{\ding{51}}}} \\

\cowalive{} 
& \textcolor{agreegreen}{\ding{51}} / \textit{\textcolor{agreegreen}{\ding{51}}} 
& \textcolor{agreegreen}{\ding{51}} / \textit{\textcolor{agreegreen}{\ding{51}}} 
& \colorbox{mismatchamber}{\textcolor{agreered}{\ding{53}} / \textit{\textcolor{agreegreen}{\ding{51}}}} 
& \textcolor{agreered}{\ding{53}} / \textcolor{agreered}{\ding{53}} 
& \textcolor{agreered}{\ding{53}} / \textcolor{agreered}{\ding{53}} \\

\grootfox{} 
& \textcolor{agreegreen}{\ding{51}} / \textit{\textcolor{agreegreen}{\ding{51}}} 
& \textcolor{agreegreen}{\ding{51}} / \textit{\textcolor{agreegreen}{\ding{51}}} 
& \colorbox{mismatchamber}{\underline{\textcolor{agreered}{\ding{53}} / \textit{\textcolor{agreegreen}{\ding{51}}}}} 
& \colorbox{mismatchamber}{\textcolor{agreegreen}{\ding{51}} / \textcolor{agreered}{\ding{53}}} 
& \textcolor{agreegreen}{\ding{51}} / \textit{\textcolor{agreegreen}{\ding{51}}} \\

\seaonion{} 
& \textcolor{agreered}{\ding{53}} / \textcolor{agreered}{\ding{53}} 
& \textcolor{agreered}{\ding{53}} / \textcolor{agreered}{\ding{53}} 
& \textcolor{agreered}{\ding{53}} / \textcolor{agreered}{\ding{53}} 
& \textcolor{agreered}{\ding{53}} / \textcolor{agreered}{\ding{53}} 
& \textcolor{agreered}{\ding{53}} / \textcolor{agreered}{\ding{53}} \\

\sticklog{} 
& \textcolor{agreegreen}{\ding{51}} / \textit{\textcolor{agreegreen}{\ding{51}}} 
& \textcolor{agreegreen}{\ding{51}} / \textit{\textcolor{agreegreen}{\ding{51}}} 
& \colorbox{mismatchamber}{\underline{\textcolor{agreered}{\ding{53}} / \textit{\textcolor{agreegreen}{\ding{51}}}}} 
& \colorbox{mismatchamber}{\underline{\textcolor{agreegreen}{\ding{51}} / \textcolor{agreered}{\ding{53}}}} 
& \colorbox{mismatchamber}{\textcolor{agreered}{\ding{53}} / \textit{\textcolor{agreegreen}{\ding{51}}}} \\

\donutweb{} 
& \colorbox{mismatchamber}{\underline{\textcolor{agreegreen}{\ding{51}} / \textcolor{agreered}{\ding{53}}}} 
& \textcolor{agreered}{\ding{53}} / \textcolor{agreered}{\ding{53}} 
& \textcolor{agreered}{\ding{53}} / \textcolor{agreered}{\ding{53}} 
& \textcolor{agreered}{\ding{53}} / \textcolor{agreered}{\ding{53}} 
& \textcolor{agreered}{\ding{53}} / \textcolor{agreered}{\ding{53}} \\

\fogalive{} 
& \textcolor{agreegreen}{\ding{51}} / \textit{\textcolor{agreegreen}{\ding{51}}} 
& \colorbox{mismatchamber}{\textcolor{agreered}{\ding{53}} / \textit{\textcolor{agreegreen}{\ding{51}}}} 
& \textcolor{agreered}{\ding{53}} / \textcolor{agreered}{\ding{53}} 
& \textcolor{agreered}{\ding{53}} / \textcolor{agreered}{\ding{53}} 
& \textcolor{agreered}{\ding{53}} / \textcolor{agreered}{\ding{53}} \\

\starship{} 
& \textcolor{agreegreen}{\ding{51}} / \textit{\textcolor{agreegreen}{\ding{51}}} 
& \textcolor{agreegreen}{\ding{51}} / \textit{\textcolor{agreegreen}{\ding{51}}} 
& \colorbox{mismatchamber}{\underline{\textcolor{agreered}{\ding{53}} / \textit{\textcolor{agreegreen}{\ding{51}}}}} 
& \colorbox{mismatchamber}{\underline{\textcolor{agreegreen}{\ding{51}} / \textcolor{agreered}{\ding{53}}}} 
& \colorbox{mismatchamber}{\underline{\textcolor{agreered}{\ding{53}} / \textit{\textcolor{agreegreen}{\ding{51}}}}} \\

\toystory{} 
& \textcolor{agreegreen}{\ding{51}} / \textit{\textcolor{agreegreen}{\ding{51}}} 
& \colorbox{mismatchamber}{\textcolor{agreered}{\ding{53}} / \textit{\textcolor{agreegreen}{\ding{51}}}} 
& \textcolor{agreered}{\ding{53}} / \textcolor{agreered}{\ding{53}} 
& \textcolor{agreered}{\ding{53}} / \textcolor{agreered}{\ding{53}} 
& \textcolor{agreered}{\ding{53}} / \textcolor{agreered}{\ding{53}} \\

\goldfish{} 
& \textcolor{agreegreen}{\ding{51}} / \textit{\textcolor{agreegreen}{\ding{51}}} 
& \textcolor{agreegreen}{\ding{51}} / \textit{\textcolor{agreegreen}{\ding{51}}} 
& \colorbox{mismatchamber}{\underline{\textcolor{agreered}{\ding{53}} / \textit{\textcolor{agreegreen}{\ding{51}}}}} 
& \textcolor{agreered}{\ding{53}} / \textcolor{agreered}{\ding{53}} 
& \textcolor{agreered}{\ding{53}} / \textcolor{agreered}{\ding{53}} \\

\pumpkins{} 
& \textcolor{agreegreen}{\ding{51}} / \textit{\textcolor{agreegreen}{\ding{51}}} 
& \textcolor{agreegreen}{\ding{51}} / \textit{\textcolor{agreegreen}{\ding{51}}} 
& \textcolor{agreered}{\ding{53}} / \textcolor{agreered}{\ding{53}} 
& \textcolor{agreered}{\ding{53}} / \textcolor{agreered}{\ding{53}} 
& \textcolor{agreered}{\ding{53}} / \textcolor{agreered}{\ding{53}} \\

\sunsalad{} 
& \textcolor{agreegreen}{\ding{51}} / \textit{\textcolor{agreegreen}{\ding{51}}} 
& \textcolor{agreegreen}{\ding{51}} / \textit{\textcolor{agreegreen}{\ding{51}}} 
& \textcolor{agreegreen}{\ding{51}} / \textit{\textcolor{agreegreen}{\ding{51}}} 
& \textcolor{agreegreen}{\ding{51}} / \textit{\textcolor{agreegreen}{\ding{51}}} 
& \textcolor{agreegreen}{\ding{51}} / \textit{\textcolor{agreegreen}{\ding{51}}} \\

\webzebra{} 
& \colorbox{mismatchamber}{\textcolor{agreered}{\ding{53}} / \textit{\textcolor{agreegreen}{\ding{51}}}} 
& \textcolor{agreered}{\ding{53}} / \textcolor{agreered}{\ding{53}} 
& \textcolor{agreered}{\ding{53}} / \textcolor{agreered}{\ding{53}} 
& \textcolor{agreered}{\ding{53}} / \textcolor{agreered}{\ding{53}} 
& \textcolor{agreered}{\ding{53}} / \textcolor{agreered}{\ding{53}} \\

\bottomrule
\multicolumn{6}{l}{\textcolor{agreegreen}{\ding{51}}: chose CPR; 
\textcolor{agreered}{\ding{53}}: chose No CPR.}\\
\multicolumn{6}{l}{\sethlcolor{mismatchamber}\hl{~~~~}: participant disagreed with \acpagent{} recommendation and chose a different CPR decision.} \\
\multicolumn{6}{l}{\underline{\sethlcolor{mismatchamber}\hl{\mbox{\hspace{1em}}}}: participant agreed with \acpagent{} recommendation despite mismatch in CPR decisions.} \\
\multicolumn{6}{l}{Unshaded cells indicate full alignment between participant and \acpagent{}.}
\end{tabular}
\label{tab:agreement-table}
\end{table*}

\FloatBarrier

Instructive feedback appeared in the post-round reflections, where participants specified additional factors for \acpagent{} to consider, such as family impact, affordability, lucidity, frailty, life stage, or systemic constraints.
These elaborations showed how participants attempted to shape not only \acpagent{}'s outputs but its underlying decision logic.

\subsubsection{Summary}
We conclude that the prototype was sufficient to simulate training an agent as an ACP proxy.
Participants were influenced by the scenarios and changed their answers across scenarios. In both agreement and disagreement cases, participants treated \acpagent{} as a system whose behaviour could be shaped over time, rather than a static decision-making tool. 
Most demonstrated corrective efforts by adjusting their preferences or giving instructive feedback to the bot after a scenario, and a small handful were less active, responding only to \acpagent{}'s recommendation and reasoning text.


\subsection{Envisioned Uses of \acpagent{}}
Building on these behavioural results, we now report the themes from our analysis of participants' experiences of training a tool when making high-risk, high-subjectivity decisions. 

In our analysis, we saw that participants' requested features drew on multiple, often concurrent framings of the role of AI in these decisions. 
Participants often shifted their framing of \acpagent{} as they worked through various concerns, including trust, uncertainty, conflict, and the need for safeguards. We believe it is likely due to the speculative, future-oriented task for a nascent, unformed agent role. 

\begin{figure*}[htbp]
    \centering
    \includegraphics[width=\linewidth]{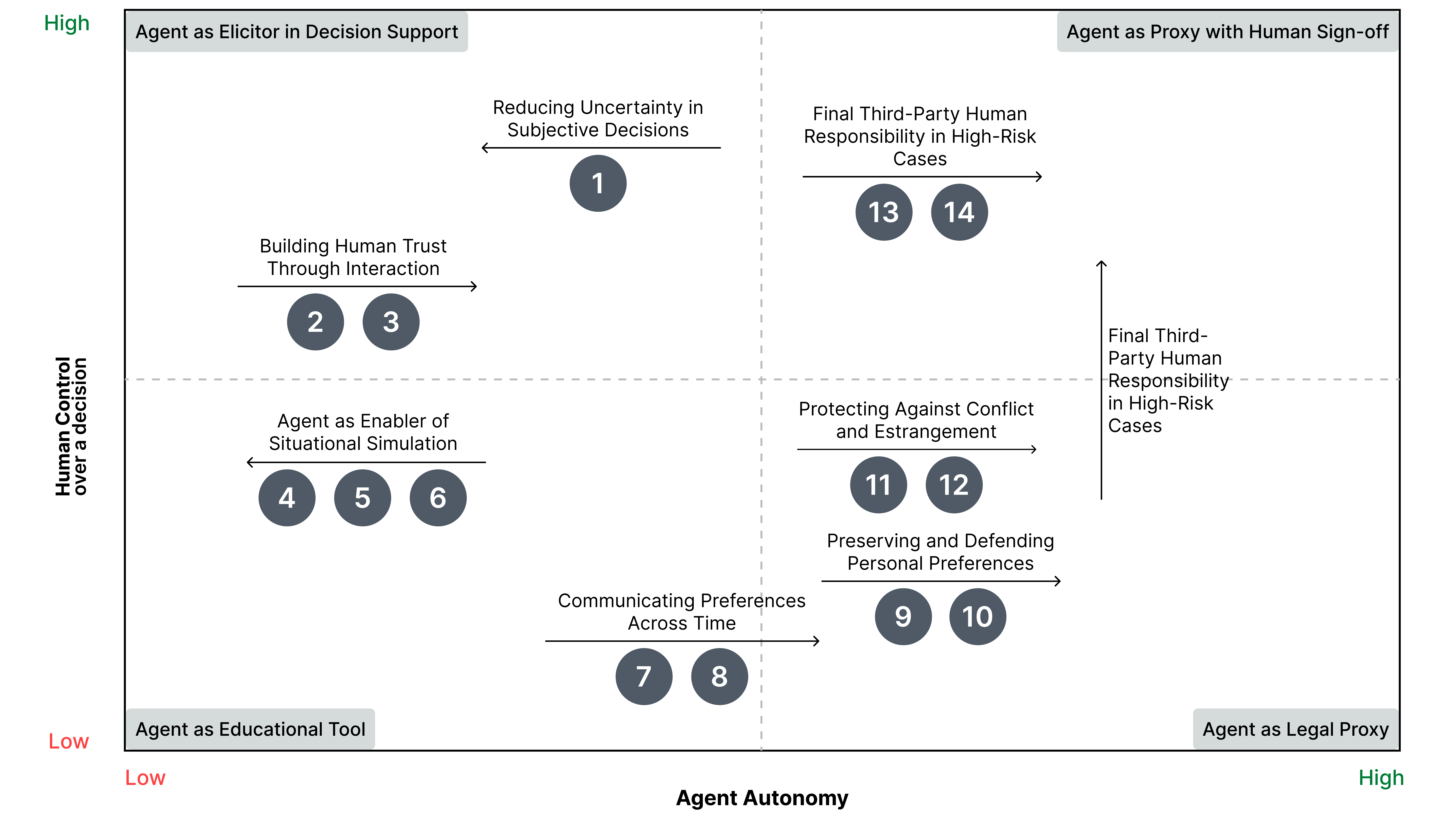}
    \caption{Quadrants of envisioned roles for \acpagent{}. 
    The x-axis represents increasing agent autonomy, while the y-axis represents human control over the decision-making process. 
    Arrows indicate pressures identified in participant reflections that pushed \acpagent{} toward or away from each quadrant. 
    Circled numbers correspond to themes presented in later subsections.
    }    
    \Description{A two-dimensional quadrant diagram mapping participant-envisioned roles for \acpagent{}. The horizontal x-axis represents increasing agent autonomy (from low on the left to high on the right). The vertical y-axis represents human control over the process of reaching a decision (from low at the bottom to high at the top).}
    \label{fig:roles-spectrum}
\end{figure*}

To best present this complexity, we mapped the themes onto \citet{shneidermanHumanCenteredAI2022}'s control-autonomy framework. The two axes yielded four broad framings of what \acpagent{} could be (see Figure~\ref{fig:roles-spectrum}). To capture the speculative aspect of the feedback, we describe the themes as 'forces' that might move the design of future proxy agents between each framing. In Figure~\ref{fig:roles-spectrum}, the numbered points are the themes, and the arrows indicate the direction of influence. 


\FloatBarrier

At lower levels of agent autonomy, participants framed \acpagent{} as an \emph{elicitor in decision support}, conversationally helping to clarify values and surface priorities, towards enabling the human user to make better decisions. 
When human control was also de-emphasised, \acpagent{} was framed as an \emph{educational tool} that could simulate difficult scenarios, helping the human to understand how decisions arose from the connection between values and decisions. Here, its value laid in preparing humans for high-risk, high-subjectivity decisions by simulating the experience thereof.

When greater agent autonomy and relatively low human control were desired, participants described requirements for a more assertive framing for a decision proxy agent: as a \emph{legal proxy} where it could advocate for or uphold prior wishes in the event of conflict or patient absence.
Conversely, there was a framing where both agent autonomy and human control were high. Here, \acpagent{} was envisioned as a \emph{proxy with human sign-off}, combining advocacy with continued oversight and ultimate responsibility remaining with a human.


In the sections that follow, we discuss each role together with the movements across the axes of human control and agent autonomy.

\input{sections/4.2.1}
\input{sections/4.2.2}
\input{sections/4.2.3}
\input{sections/4.2.4}

%% file: sections/4.2.1.tex
\subsubsection{\acpagent{} as an Elicitor in Decision Support}
In this quadrant, \acpagent{} was envisioned as an engaging way to reduce uncertainties and elicit preferences (\leavevmode\Circled{1}). 
By contrast, participants also noted that more agent autonomy might be acceptable from the decision support tool if there were more trust-building interactions with the agent (\leavevmode\Circled{2}, \leavevmode\Circled{3}).

\paragraph{\leavevmode\Circled{1} Reducing Uncertainty in Subjective Decisions: Add Decision Parameters.}

Participants struggled with the inherent uncertainty of subjective, preference-sensitive decisions. To cope, they requested more detailed, personalised information. The hope seemed to be that with more certainty in the data, it would help them resolve the decisional dilemma. 

Firstly, participants wanted scenarios to have more details of diseases. These were details that personally mattered to them and that they knew would change their decision. 
\pumpkins{} requested more alternatives and details on complications, side effects, and treatment options. They noted that dementia or semi-conscious states were not represented, even though \acpagent{} generally \dquote{just followed what I want [...] never said no, it said yes.}

\trombone{} similarly asked for very specific distinctions in scenarios:
\begin{quote}
    \dquote{Like what kind of stroke? [...] bedridden can be only like you need not stay on the bed all the time even though you are bedridden [...] The scenarios may be more well-defined, more detailed.}
\end{quote}
Participants also wanted to know what their own response to the illness might be, requesting details that would concretise the specific situation they might find themselves in.
\goldfish{}, who rated pain tolerance 3/10, explained: 
\begin{quote}
    \dquote{I would like the bot to specify how much suffering in the scenarios [...] because I don't know how much suffering I'm supposed to be in in that scenario [...] it's very hard to imagine [...] so the bot should describe the symptoms and the suffering so I can decide. Otherwise, I'm just going to say yes, give me CPR, because I'm assuming I return to normal [...] If I'm going to end up with a tube in ICU (Intensive Care Unit) until I die, then of course I'm going to say no. So the bot should point those things out, and what the chances are.}
\end{quote}
Building on the above, \goldfish{} suggested that a decision-support agent could ask probing questions such as, \dquote{What if you were suffering in that condition?}, or \dquote{Would you still want whatever [situation] your mom was in that condition when she had?}

Secondly, the request for data extended to concerns beyond the clinical situation. For example, several participants called for more explicit assessment of financial outcomes.
Although the trade-off question was phrased in terms of personal financial assessment, this did not provide enough detail. 
\redpasta{} suggested quantifiable thresholds (e.g., \dquote{\$500,000}) to avoid burdening her kids, while \toystory{} in another session called for surfacing subsidies and social assistance: \dquote{Not everyone is well-to-do, so [...] the social assistance they can get [...] that component would be good also.} 

To further enhance the decision support elements, participants proposed integrating personal financial information. \sticklog{} linked finances to inequities in pain relief provision, reflecting that affordability could shape not only access to treatment but also the degree of comfort afforded in long-term care.
To further enhance these non-clinical decision support elements, participants proposed integrating personal information. (\sticklog{}): \dquote{If you could find a way to input your own financial data or measurements for pain tolerance, then I think the results would be much better.} 
Some wanted very specific, personalised financial information that may not even exist in databases. For example, \donutweb{} asked a question about finance for the family as a whole:
\begin{quote}
    \dquote{Can I really afford? Can my family really afford? And do they want to do it? I've heard of stories where the [insurance] all drain and there's a lot of stress.} 
\end{quote}
We found that these requests (individualised prognosis, unknown physical pain responses, localised cost information) are good-to-have for decision-making, but questions remain about how feasible obtaining such information might be.
They require strong predictive models, may use data outside current health record databases, and may even be unknowable (e.g. physical pain response) \cite{fooBenefitsRisksLLMs2025}. 
However, if these decision parameters could be integrated, then a proxy agent system would be useful as a decision-support system with humans making the final decision.

\paragraph{\leavevmode\Circled{2} Building Human Trust Through Interaction: Adjust Presentation to Align with Decision Style.}
Participants also described conditions where enriching the interaction style and increasing the representation of decisional uncertainty would make \acpagent{} more reliable and supportive as a decision tool. 
Several found binary yes/no choices inadequate and wanted to more carefully bound their choice, with \donutweb{} remarking: \dquote{The yes-no puts me in a difficult spot because [...] I was thinking, yes, maybe with some condition. No, with some condition.}  

Others called for a more conversational tone: \starship{} expected \dquote{a human figure you’re talking to instead of just purely words,} while \fogalive{} preferred phrasing that felt \dquote{like talking to you,} and \toystory{} suggested softer phrasing to reduce the \dquote{bot-likeness}.
\toystory{} posited that phrasing such as \dquote{have opted} or \dquote{recommend or not recommend} felt jarring, preferring a more natural conversational flow. 

The current presentation of results in text form was also an issue. 
\webzebra{} proposed new ways of visualising the binary decision, suggesting percentages or graphics to avoid a \dquote{judgmental} yes/no: 
\begin{quote}
    \dquote{Let’s say 60\% is a yes, 40\% no [...] rather than just a yes or no, because it sounds like gacha\footnote{Referring to a gambling game} yes-no, very judgmental. But graphical form would be a good representation, rather than just reading, because we get lost then.}
\end{quote}
He described the current \dquote{huge wall of text} as off-putting, and suggested visual elements such as a status bar or pie chart.

\paragraph{\leavevmode\Circled{3} Building Human Trust Through Interaction: Provide Details of Provenance.}
Desires to improve security, governance, and provenance pointed towards a future in which an agent proxy could take on a somewhat more autonomous role, provided these safeguards were in place.
Participants expressed openness to relying more strongly on \acpagent{} if it were embedded in trusted infrastructures or transparent institutional arrangements.
\grootfox{} worried about data protection, cautioning that \dquote{individual[s] don't want to input so many (sic) personal information [...] because you have no way to control to safeguard [...] the information will not be hacked one day.} 
\sticklog{} similarly stressed that trust depended on governance: \dquote{It depends on the conflict of interest of who's running the bot, who owns the bot.}  

Others suggested that integration with secure infrastructures could build confidence. 
\toystory{} proposed linking ACP records to national systems so only the individual could access them: \dquote{If I'm using [it] to enter my ACP [...] only I can access that. Then I would be okay with that. Because I want to know [...] how much am I covered.}  

In the group discussion, \donutweb{} added that these integrations would help surface practical cost implications that would help him make more informed choices: \dquote{How many more days I can stay? Yeah, it's not cheap. Even community hospital is not cheap.}
To which, \toystory{} agreed that decisions should be \dquote{based on what I have}, while \fogalive{} concluded that such integration would be \dquote{very practical}.

Finally, \sticklog{}, in another session, also highlighted the need to integrate an agent proxy with existing government instruments: 
\begin{quote}
    \dquote{If the ACP is aligned with all the other legacy planning tools, then it might be more useful [...] whether it's ACP, LPA [...] will, funeral plan, so on and so forth. And then moving forward, [...] assisted suicide [...] can become a composition at some point. And maybe the bot will help in those scenarios.}
\end{quote}
However, \cowalive{} cautioned that, regardless of provenance, there was a concern with drift and a single point of acceptance. Instead, she suggested that trust would increase with more reliable interaction, recounting that \acpagent{} once suggested CPR when she had indicated to refuse CPR.
She emphasised that trust depended on repeated testing: \dquote{My confidence in the bot will depend on how frequently I test it, how deeply I tested and also how recently I tested it.}  

%% file: sections/4.2.2.tex
\subsubsection{\acpagent{} as an Educational Tool}
In this quadrant, participants desired to develop the agent learning, reflection, and value clarification, shifting the framing and reducing the relevance of both human control and agent autonomy. 
Rather, the role of the agent should be to enable learning about high-risk, high-subjectivity decisions through rehearsing scenarios and prompting self-reflection without directly influencing real outcomes.
Specifically, participants wanted to rehearse possible situations and reflect on their own values (\leavevmode\Circled{4}) or simplify otherwise complex ACP processes (\leavevmode\Circled{5}). 
Furthermore, because some valued \acpagent{}'s ability to guide reflection and surface their priorities, they also expanded its use, such as for facilitating difficult family conversations (\leavevmode\Circled{6}).
The ability of the agent to guide learning and record their responses led the participants to further imagine that it could act as an advocate (\leavevmode\Circled{7}, \leavevmode\Circled{8}) when it was given the role of reproducing their responses.


\paragraph{\leavevmode\Circled{4} Enabling Situational Simulation: Bidirectional Training.}
Participants described the interaction with \acpagent{} as reciprocal or bidirectional: sometimes empowering them to shape \acpagent{}, other times revealing its influence on their own position.

\baritone{} described the experience of training the agent in this way: \dquote{At the end of the whole experience, it felt more like I was training the bot on the [ACP] thing,} and emphasised the value of the training process for prompting reflection: \dquote{I think it's good to do this exercise, to make you reflect [...] going through the exercise may take you through a different type of thing that will happen.}  

Similarly, \seaonion{} indicated that \acpagent{} might help her to learn more about herself.
\begin{quote}
    \dquote{When I put in my preferences [...] some of the recommendation[s] [...] really [are] what I want [...] I don’t know how to speak for myself but the bot give (sic) me words that describe what I’m feeling.}  
\end{quote}

\redpasta{}'s response suggested that it was the prose-like response that helped bring disparate ideas together: \dquote{Once I gave my priority or outlook of life, it managed to capture that. So the subsequent scenario recommendation is very aligned with what I choose.}  

Others highlighted how \acpagent{} was also training them to reflect on ACP.
\cowalive{} explained:
\begin{quote}
    \dquote{It was not so much me training the bot, I felt the bot was training me to do the ACP properly [...] The questions asked so far seem to be triggering the right thoughts about what I should tell the bot.} 
\end{quote}
\cowalive{} described referring back to the CPR notes and adjusting their preferences, viewing the prompts as \dquote{useful [...] because I get a scenario, I think, I refer, and then I make my decision, and then I get to adjust again my preferences.} 

Newcomers to ACP particularly seemed to value this low-risk space for reflection on high-risk situations.  
\starship{} said:
\begin{quote}
    \dquote{This is my first time hearing ACP [...] Using AI [...] to input all this is rather new to me [...] I find the questions are quite challenging but certain scenarios really hit me.}  
\end{quote}
Their comment prompted \fogalive{} to note: \dquote{Doing ACP, setup will, LPA, all these are important [...] Only if a person goes through a situation like that, then you think of what to do.}  
\starship{} further described a recent hospitalisation that forced reflection on unfinished business, the desire to spend time with loved ones, and fears of burdening their family without an LPA: \dquote{The first thing that came to my mind was [...] what [would] happen to my family?}  
For him, engaging with a simulation-focused \acpagent{} could help him explore the practical and emotional ramifications of these decisions.

\paragraph{\leavevmode\Circled{5} Enabling Situational Simulation: Simplify ACP Processes.}
Several contrasted this interactive reflection with their knowledge of existing ACP processes.
\fogalive{} shared, \dquote{Of all things that I've done, I've not done ACP [...] when I went into the website, there were so many questions until I don't have guidance [...] But if this chatbot is interactive enough I think it will help me a lot.}  

\toystory{} similarly found that early indications of their preferences allowed her to keep better track of what she prioritised: \dquote{I liked that in the beginning already, I could [...] take stock of what are the things. So it was simple and straight to the point.}  

In contrast to this support for \acpagent{} as a \squote{trainer}, \grootfox{} cautioned that without human professional facilitation, the agent risked falling short.
Drawing on his knowledge of Taiwan’s model of nurse-led ACP, he argued that an AI-supported agent alone could not provide trained guidance to make informed choices and would be inadequate for guiding such complex and sensitive choices.
Additionally, \sunsalad{} criticised \acpagent{} for being too agreeable, explaining that if it had challenged their inputs or helped to break down complex scenarios, it could have helped him to learn more:
\begin{quote}
    \dquote{Perhaps you can input it to be a bit more critical [...] It's just very agreeable [...] this bot doesn't help us to then break down the scenario so it just kind of follow through.}  
\end{quote}
  
\paragraph{\leavevmode\Circled{6} Enabling Situational Simulation: Facilitate  Family Conversations.}
Having found value in \acpagent{} as a tool for reflection on values, some participants began to imagine \acpagent{} facilitating conversations for someone else in their family, despite being asked to use \acpagent{} for themselves during the study.
This was most evident in Workshop 3, where several participants were already caregiving for their parents, even though none had previously acted as a Nominated Healthcare Spokesperson or LPA donee. 
They saw potential for an agent to serve as a facilitator of difficult family conversations.  
\toystory{} described how their own reflection immediately extended to their mother: 
\begin{quote}
    \dquote{For me, when I'm doing this [...] I think about myself, but more importantly, I also was thinking about my mother [...] I want to have this conversation with her because it's not really easy to have this kind of conversation.} 
\end{quote}
\toystory{} noted how language and navigation barriers often left her acting as an interpreter during medical appointments, and hoped the tool could allow their mother to document her own wishes: 
\begin{quote}
    \dquote{I want it from my mom's own mind [...] I don't want to feel like guilt or anything of having made the wrong decision when that's not what she wants.} 
\end{quote}
Furthermore, if the system were accessible at primary or community care level, \toystory{} proposed it could encourage earlier and more inclusive ACP discussions, especially if caregivers were invited to walk through it together with their care recipients. 

\starship{} noted cultural taboos around discussing death: \dquote{My parents are all in their 80s. So when they say, what is this ACP? Then they say, hey, I'm still alive. The talk of death is so sensitive.}
\fogalive{} echoed this generational divide: \dquote{I’m glad that starting our generation onwards, we are quite open in this area. But then for a lot of seniors, they’re still very reluctant to talk about this.}  
\fogalive{} suggested that an interactive design might help elders overcome reluctance: \dquote{Like talking to a friend like that. Like talking to a friend but it's a robot.}  

The ability of \acpagent{} to record elicited preferences suggested to participants that it might be developed to help users communicate their preferences over time. In \ref{fig:roles-spectrum}, these themes create pressures away from education alone, towards framing the agent as a proxy. 
 
\paragraph{\leavevmode\Circled{7} Communicating Preferences over Time: Capture Preferences While Lucid.}
Participants stressed the agent might enable them to capture preferences at their most lucid stage, before crises struck, similar to the current use of ACP.  
\donutweb{}, although new to ACP, suggested, \dquote{Maybe another entry point would be when [...] something happened but they're still sound before things turn for the worse.}
\toystory{} emphasised the need for advance planning: \dquote{I don't want to be in that state of mind when I'm doing this. I want to do it in advance when things are not drastic.} 
\toystory{} added that sudden accidents also made timing unpredictable: \dquote{It may not all be illness related [...] It might be something else that can happen to you and then you are put in that kind of position suddenly.} 

Others grounded their reflections in medical trauma, highlighting that well-captured preferences in \acpagent{} could serve as a reminder and reinforcement of earlier preferences, reducing confusion at the most critical moments.
\webzebra{} recounted how, in his mother’s final hours, panic led him to contradict her stated wish: 
\begin{quote}
    \dquote{My mum [...] said she wanted a quiet death. But everything was too crazy [...] I mistold the doctor. [...] But it wasn’t the way she wanted to go.} 
\end{quote}

These findings suggest that our participants did not account for how the clinical situation may change over the course of illness and remains unpredictable. As with current ACP processes, decisions made in the present may not align with future circumstances, raising similar challenges to existing practices in terms of ensuring that current documentation remains relevant as values evolve.

Just a small handful were aware of potential clinical uncertainties:
\trombone{} and \grootfox{} emphasised that technology and medical treatments evolve quickly, making current "terminal" prognoses or resuscitation outcomes potentially outdated in months or years.
This underscored the uncertainty of making decisions too early and the subsequent need for periodic review.

\paragraph{\leavevmode\Circled{8} Communicating Preferences over Time: Adapt to Life Stage.}
Participants' reflections on how age and life stage shaped their decisions raised interesting questions about how an agent might serve as repository of preferences over time.
\seaonion{} stressed that values could shift over time, noting that financial and pain considerations might matter differently over time and that preferences should be preserved when most lucid: 
\begin{quote}
    \dquote{Maybe certain seasons of my life, there’s a cost concern, but maybe certain seasons [...] cost is not a concern. Or the degree of pain, it depends what kind of sickness. Maybe I got dementia, I don’t even know my pain.}
\end{quote}

Younger parents emphasised their responsibilities to children, while older participants spoke of readiness to \dquote{let nature take its course.}
\redpasta{} contrasted the needs of younger parents (e.g., \baritone{}) with those whose children were grown:
\begin{quote}
    \dquote{I think the age matters a lot [...] Actually for me, I'm quite happy now that I've crossed the [...] half of the century, right? So if really things happen, I'm quite well prepared already. My kids are all grown up and stuff. I got no [...] so-called things that I cannot let go. So to me, I kind of let nature take its course.}
\end{quote}
Following this, \redpasta{} later added, \dquote{the mentality is different at a different stage of life. Certain things 你会看得很开  (you will be able to see it with an open mind), certain things you will cling on very hard, like to [\baritone{}'s] point, her kids are still young.}

\trombone{} suggested that life stage might matter more than clinical change:
\begin{quote}
    \dquote{My friend [...] as long as she's not terminal and she's not unconscious she'll want to keep on going [...] but medical science is advancing. Six months down the road or even one year new drugs may come out. [...] So this thing really has to define at which point you want to let go. Like for me, just like I said, if I'm 88, I don't care. I just want to leave, waste all my money. That's it.}
\end{quote}

This discussion of change led to \redpasta{} being uncertain on whether \acpagent{} should remain a fixed directive or evolve alongside her, expressing a desire for the system to be a \dquote{buddy} that could accompany her through life stages: \dquote{If it's going to serve as [...] a buddy to me [...] when I'm age 60 [...] then yes, those features are important. Because as you age [...] your priority of life is different, and at the stage of your different family members, you also have different considerations.}  
Yet \redpasta{} also worried about the risks of continual learning, still preferring the stability of static instruments such as LPAs or Advance Medical Directives (AMDs).

%% file: sections/4.2.3.tex
\subsubsection{\acpagent{} as a Legal Proxy}
Beyond the use of \acpagent{} for static current wishes, some participants were extending the autonomy of \acpagent{} to the point of actively influencing and, further, dictating the final decision.
Here, participants expressed a desire for \acpagent{} to not just preserve their wishes, but potentially advocate for them, in order to protect against conflict or estrangement.
In envisioning higher autonomy for \acpagent{}, some participants wanted it to take a more active stance, representing or "fighting for" their preferences in contested moments (\leavevmode\Circled{9}) and ensuring that lucidly expressed wishes (see \leavevmode\Circled{7} above) would be preserved even if family members objected (\leavevmode\Circled{10}).
For similar reasons, participants also hoped a legal proxy could reduce the emotional burden on families, especially in situations where guilt, conflict, or estrangement might arise (\leavevmode\Circled{11}, \leavevmode\Circled{12}).
This positioned \acpagent{} as a neutral arbiter that could protect both patients and loved ones from the psychological toll of decision-making.

\paragraph{\leavevmode\Circled{9} Preserving and Defending Personal Preferences: Take an Active Role in Decision-Making.}
\goldfish{}, who had experienced heart failure and chronic lung disease in others, explained that she would prefer \acpagent{}’s authority over doctors who may be legally compelled to perform CPR: 
\begin{quote}
    \dquote{Better than no decision [...] Sometimes better than your own relatives that want you to live no matter what. [...] I don't think I would trust the doctor to decide. I would prefer the bot. The doctor is required by law to perform CPR, so definitely he will do CPR.} 
\end{quote}

\paragraph{\leavevmode\Circled{10} Preserving and Defending Personal Preferences: Reinforce Wishes.}
\fogalive{} explained that lucid wishes should override family emotions: 
\begin{quote}
    \dquote{If I have really done this based on my personal preferences, when I am still clear-minded, then I think there shouldn’t be too much emotions coming in [...] I hope the doctor will respect my wish rather than prolonging my life.} 
\end{quote}
\sunsalad{} imagined \acpagent{} reasoning in their absence: \dquote{So it will be an intelligence with the database to make the wisest decision possible at that time already. No one is around, right? Or help the doctor, perhaps.}

\paragraph{\leavevmode\Circled{11} / \leavevmode\Circled{12} Protecting Against Conflict and Estrangement: Family Dynamics / Decisional Burden.} 
\baritone{} reflected on their husband’s death and her hope that \acpagent{} might relieve their children of guilt over not having tried to save her:
\begin{quote}
    \dquote{In this case, I choose CPR for most of the situations because I know CPR is not very effective. So do [CPR] just to make everyone happy, just to know they did something [...] Yeah, they tried. I think the important thing is they tried.}
\end{quote}
\baritone{} stressed that her preference should be followed, because her principle was that the emotional effect on loved ones outweighed her own outcome: \dquote{I die doesn’t matter. The [ones left behind are] the one[s] guilty.}
In the group discussion, \redpasta{} suggested that \acpagent{} could act as \dquote{an independent assessor or recommender} so that difficult decisions would not fall directly on \baritone{}’s children’s shoulders.

%% file: sections/4.2.4.tex
\subsubsection{\acpagent{} as a Proxy with Human Sign-off}
In this quadrant, participants envisioned \acpagent{} as operating with strong autonomy, capable of representing, advocating for, or interpreting their preferences, yet always under conditions of high human control.
This quadrant reflects attempts to balance trust in agent autonomy with the enduring need for human responsibility in high-risk decisions.
Even when participants trusted \acpagent{}, some emphasised that responsibility for high-risk, high-subjectivity decisions should always remain with humans (\leavevmode\Circled{13}). 
At the same time, some wanted \acpagent{} to counteract emotionally clouded human judgement without overriding human authority (\leavevmode\Circled{14}).
Together, these forces positioned \acpagent{} as a legally active proxy that could fight for patient wishes with increased autonomy, but always under human sign-off and control.

\paragraph{\leavevmode\Circled{13} Requiring Final Third-Party Human Responsibility: Use Human Witness as a Safeguard.}
Some participants saw value in pairing \acpagent{} with a human witness, mirroring existing ACP processes. 
\cowalive{} explained: \dquote{I do trust it because I trained it [...] But I also want somebody to [...] see that I’ve done this process. It’s just so cold when it’s just me and the bot.} 

It was important for a human third party (i.e., clinicians, family members) to be involved in the decision.
\trombone{}, who agreed with \acpagent{} in three out of five scenarios during the agent-training stage, explained: \dquote{As of now, I can only trust the doctor. The only subject matter expert.}  

\donutweb{} emphasised that even when ACP records are in place, families would still want to be consulted for closure: \dquote{If we have done ACP beforehand, then the doctor [...] No, they will still ask. Because your family member might [...] want to have a proper closure.}
\sticklog{} similarly believed decisions about life support could never be outsourced: \dquote{It’s a very painful decision [...] so the bot cannot make it,} despite agreement with \acpagent{} in four of the scenarios. 

\sticklog{}, \donutweb{}, and \toystory{} stressed that the agent lacked empathy, with \sticklog{} suggesting that families might think it would be better to seek advice from healthcare professionals than from a \dquote{robot}.  

\paragraph{\leavevmode\Circled{14} Requiring Final Third-Party Human Responsibility: Support through Neutrality.} 
In contrast, \fogalive{} and \sunsalad{} saw neutrality as a safeguard against emotionally clouded human judgments. 
\sunsalad{} argued:
\begin{quote}
    \dquote{The AI should be unbiased, should remove the emotion part. That means you cannot put your personal experience into it, only because there can be different possibility. So there's no right and wrong in such situation[s], so the AI give [their suggestion] and then [let] the person decide.}  
\end{quote}
\toystory{}, despite noting that end-of-life decisions were emotionally laden, emphasised that it was also important to balance objective and non-emotional decisions since ACP was a \dquote{heavy topic}.
She further shared that the simplicity of \acpagent{} and the agent-training stage \dquote{was very factual [and] straight to the point.}

%% file: sections/5_discussion.tex
\section{Discussion}
This study revealed both opportunities and limits of introducing an AI proxy into ACP, not only raising questions of trust, delegation, and authorship, but also the broader problem of how to design for such inherently high-risk, high-subjectivity medical decisions.
We discuss our findings as such: (1) what we learnt about using AI in such contexts; (2) viability of the envisioned roles; and (3) design implications for future ACP systems.

\subsection{What We Learned About Using AI in High-Risk, High-Subjectivity Decisions}
\rrev{8}{kel}{expanded the critical reflection on appropriateness and limits of AI/LLM involvement in high-subjectivity, high-risk ACP decisions, incorporating reviewer-suggested concerns (e.g., relational benefits of human ACP, responsibility gaps, and "not to design" considerations). Clarified appropriate use cases and boundaries of ACPAgent.}
Our findings suggest that the highly subjective nature of ACP decisions may lead to consistent disappointment in proxy agents, particularly when framed as decision-support tools.
We found participants were somewhat aware of the subjective nature of the decisions, resisting recommendations when presented as dichotomous yes-no answers. 

They suggested an \textit{interactional style} that allowed them to incorporate the elements of uncertainty and subjectivity that they deemed relevant, including personal conditions, varying tolerance for suffering, and contextual factors like prognosis, cost, and family circumstances. Furthermore, each decisional element mattered differently to each participant, underscoring how personal these decisions are. 
Participants also valued proxy agents' conversational tone, reflective prose, and bi-directional training, which supported reflection about these decisions. Some even felt that they were being taught how to make decisions and how those decisions might map to their stated preferences. This educational value extended to supporting fraught conversations with others. 

Yet high subjectivity decisions also created unique challenges, especially when decisions are subjective to the point of being preference-sensitive.
Alongside more feasible requests, such as details of the disease, participants occasionally requested additional details that were not feasible, including personalised forecasts of pain, the cost of care to the patient and their family, and how long the savings might last.
Such information falls under the umbrella of what \citet{epsteinDevelopmentAdvanceCare2017} calls ‘epistemic uncertainty’ in end-of-life care decisions, which includes data such as prognoses and unpredictable changes in the patient’s clinical situation. 
These data are not only hard to obtain but may also be unknowable.
Our findings suggest that, unlike human-only ACP processes, participants seemed to expect that technology, computational tools and ‘AI’ might be able to help reduce the challenge of subjective decisions by supplying this type of data. 

Thus, particularly when framed as a decision-support tool, AI-enhanced decision support must \textbf{moderate expectations for objective data or be accompanied by more education about the nature of highly subjective decisions}. In planning for future care decisions, the need for trade-offs in serious illness means that decisions often feel unsatisfactory. For example, decisions might require trade-offs between quality of life and time gained.
Nonetheless, some participants approached the scenarios as if there were a hidden "right" choice, and when none was forthcoming, wanted more detail on prognosis, suffering, and cost. The nexus of high risk, high subjectivity and lack of awareness of the need for trade-offs suggests that far more education and awareness must precede the use of an AI agent for future care planning. Additionally, just as with patient decision aids~\cite{staceyDecisionAidsPeople2017}, they should \textbf{permit and encourage different weightings} of the decisional elements. 

Given the high-risk nature of the decisions, our findings suggest that \textbf{trust-building remains central to AI system design}. As with previous research in trust~\cite{ashooriAIWeTrust2019}, our study also showed the influence of interactional style, provenance, and system reliability over time. 
However, our study provides evidence that there is interest in utilising proxy agents to counter this high-risk nature by using them as safe spaces to explore difficult decisions while having the legal proxy make decisions that protect loved ones from bearing guilt or regret.  
Additionally, \acpagent{} was perceived \textbf{as a neutral party, having no stakes in any outcome}, thereby less likely to bias outcomes in any particular direction. 

This echoes \citet{candrianRiseMachinesDelegating2022}, who showed evidence of an increased willingness to delegate to machines when decisions might involve personal loss and is consistent with \citet{steffelPassingBuckDelegating2016}, who showed that people are more likely to delegate choices to others, especially when negative consequences exist. This may be an effort to avoid blame or feeling responsible for such decisions, thereby raising risks of using agents as legal proxies, such as moral de-skilling over time~\cite{vallorMoralDeskillingUpskilling2015} and questions of ethical agency. 
Furthermore, greater agent autonomy might also reinforce negative outcomes of agents that might be manipulated during training towards certain decisions, particularly when the source of manipulation is unclear.
Risks of manipulation underscore the need to audit and safeguard the training process to prevent malicious intent.

\subsection{Feasibility of the Framings of \acpagent{}}
\begin{figure*}[htbp]
    \centering
    \includegraphics[width=\linewidth]{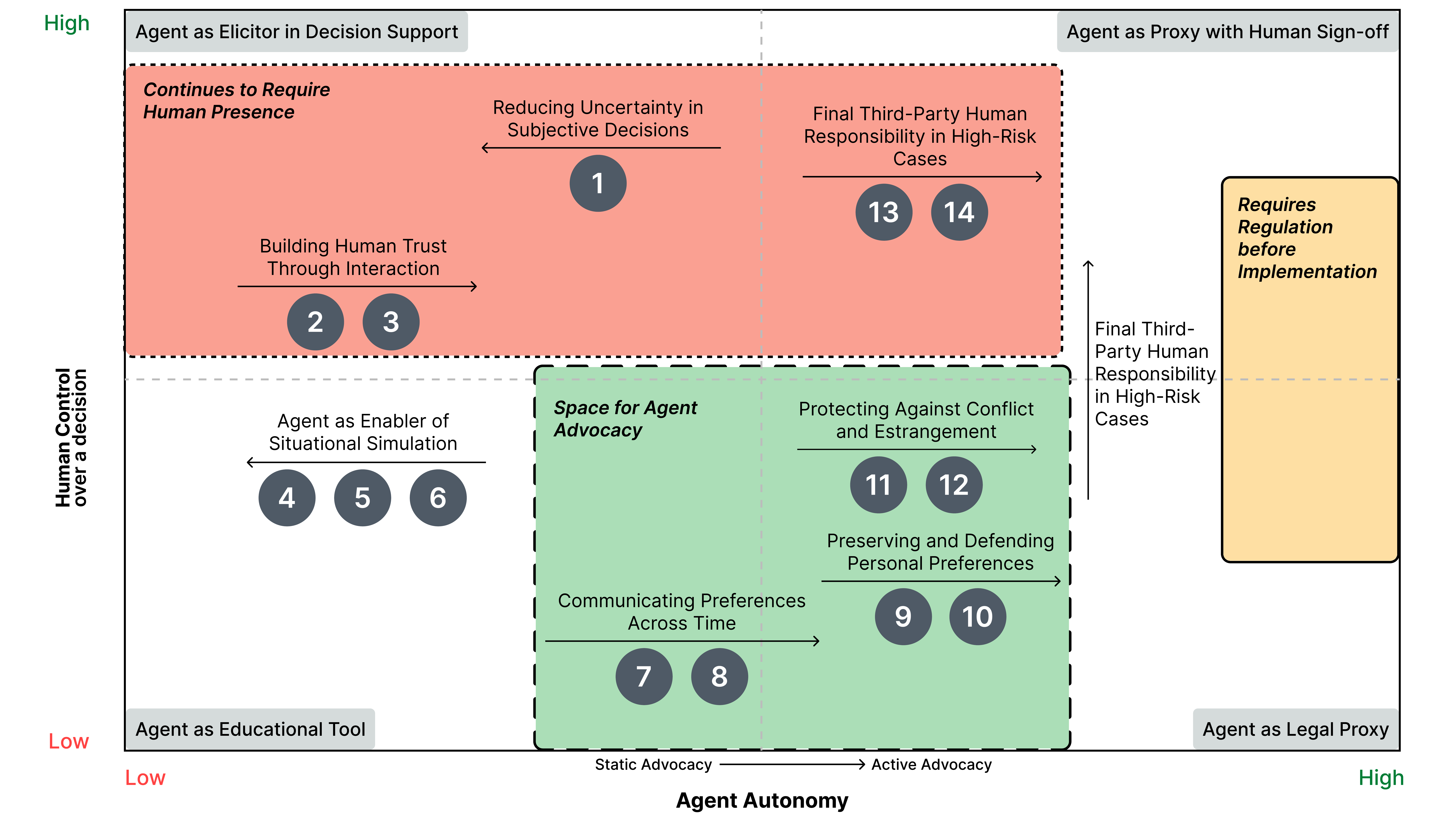}
    \caption{\textit{Framings of \acpagent{}} situated within a two-dimensional space. 
    The x-axis represents agent autonomy, and the y-axis represents human control over the decision-making process. 
    Quadrants capture distinct framings observed in our study, while the middle shaded region highlights advocacy as an emergent framing. 
    The other two shaded regions (pink and yellow) refer to areas that face systemic constraints: those requiring high human control (limited by caregiver availability) and those dependent on regulatory frameworks (plausible only with government authorisation).
    Circled numbers correspond to themes discussed in the earlier Results sections.}
    \Description{Quadrant diagram with agent autonomy on the x-axis and human control on the y-axis. Four quadrants are labelled: decision-support tool, educational tool, proxy with human sign-off, and legal proxy. A green central shaded region (with dotted boundary lines) highlights "advocacy" as an emergent middle ground. Two other shaded boxes mark framings facing systemic constraints: those requiring high human control (limited by caregiver availability), shaded in pink (with dashed boundary lines), and those dependent on regulatory frameworks (plausible only with government authorisation), shaded in yellow (with solid boundary lines). Circled numbers map to themes from the earlier Results section.}
    \label{fig:roles-discussion}
\end{figure*}

Given the discussion above, we now consider the feasibility of each framing that we observed. Participants most often positioned \acpagent{} as a decision-support tool or a proxy (see Figure~\ref{fig:roles-discussion}). 
Here, however, we interpret these results through the lens of feasibility, represented by the orange and green boxes in the figure. The lens of feasibility considers what is plausible, problematic, or requires systemic change.

\rrev{4}{kel}{Revised Table 2 and Figure 10 for accessibility and black-and-white readability per 2AC, adding clearer legends, redundant cues, and distinguishable boundary markers.}

\subsubsection{Framings Requiring High Human Control}
Framings with high human control, namely decision-support tools and proxies with human sign-off, still rely on the presence of human caregivers or professionals to witness, validate, or interpret decisions.
Although human oversight remains desirable in high-risk decisions, demographic shifts in ageing nations may require exploring solutions that do not solely depend on the presence of a trusted other, such as a close family member or a healthcare worker. 

\subsubsection{Framings Dependent on Regulatory Frameworks}
Imagining \acpagent{} as a legal proxy or one with human sign-off surfaces issues of law and governance. 
Such roles are feasible only if regulatory frameworks evolve to recognise agents as legally authorised proxies, contingent on government adoption and professional consensus.

\subsubsection{Advocacy as a Middle Ground}
Among the framings, \emph{advocacy} appeared the most feasible and desirable option in the medium term of 5-10 years. 
\textbf{This advocacy framing adds a novel role for AI in future care planning, moving beyond previously identified uses for passive extraction and prediction} \cite{ariozArtificialIntelligencebasedApproaches2025} to a more active agentic role. Situated between tool and proxy, it positions \acpagent{} as a bounded representative of patient wishes: able to amplify and reinforce elicited values, yet still variably complementing human oversight rather than displacing it. 
For agentic advocacy to be more feasible, however, our findings suggest that three issues must be addressed:
\begin{enumerate}
    \item Educating users to challenge the false belief that preferences remain stable~\cite{zivkovicForecastingForeclosingFutures2018}.
    \item Clarifying when agent advocacy should remain \textit{static (replaying preferences without adaptation)} or become \textit{active} (arguing on behalf of the patient, with the possibility of evolving alongside them or serving as an emotional buffer)~\cite{candrianRiseMachinesDelegating2022}.
    \item Defining the degree of authority such advocacy should hold relative to families and clinicians in shared decision-making, an issue closely tied to the parameters of \textit{provenance} (who made it?) and \textit{deployment timeline} (when does it start and end?) in generative ghost design~\cite{morrisGenerativeGhostsAnticipating2025, sediniAdvanceCarePlanning2022, suExperiencesPerspectivesFamily2020}. 
\end{enumerate}

Some participants pushed this vision further, imagining 
\acpagent{} as a long-term companion that could \dquote{grow over time} with its user (e.g., as with \redpasta{}'s \dquote{buddy} description.)
Yet even here, participants worried about the risks of continual adaptation, favouring the known stability of static instruments such as LPAs or AMDs.  
Further research is needed to investigate how a system should account for changes associated with its trainer's age and health status.
This tension underscored competing desires for flexibility versus certainty in ACP, echoing broader work on deliberative exchanges with AI as both externalising reflection and prompting self-correction~\cite{jeonLettersFutureSelf2025, maHumanAIDeliberationDesign2025a}.

\subsection{Addressing Limitations of Current ACP and Design Considerations for Future Care Planning Tools}
\rrev{3}{kel}{revised contribution framing and strengthened articulation of limitations in current ACP processes in response to 2AC. aligned introduction and discussion so the design contribution is clearly motivated and reflects the scope of Section 5.3 (which has also been refined).}
Building on the above, we further consider whether this advocate agent framing can meaningfully address limitations in ACP and what this suggests for future care planning tools more broadly.
As we outlined earlier, existing practices face three widely recognised constraints.
First, ACP decisions are infrequently revisited, despite long-standing calls to treat ACP as a conversational process that evolves with changing circumstances~\cite{hickmanCarePlanningUmbrella2023, sudoreDefiningAdvanceCare2017, morrisonWhatsWrongAdvance2021}.
Second, ACP materials are frequently difficult to retrieve at the point of care, creating gaps between documented preferences and real-time decision-making~\cite{malhotraComplexityImplementingNationwide2022, kuusistoAdvanceCarePlanning2020}.
Third, many individuals struggle with linking broad values to specific medical decisions, especially in high-risk and high-subjectivity contexts such as CPR~\cite{epsteinDevelopmentAdvanceCare2017, morrisonWhatsWrongAdvance2021}.

To address gaps in translating personal values into concrete treatment decisions, scenario-based reasoning and more explicit rationales can help users see how their priorities translate into specific outcomes.
With 77.3\% of CPR decisions in our study being shaped by such details, narrative framing could provide a meaningful scaffold for reflection, rather than a hidden mechanism of influence.
This aligns with work showing that vignettes and contextual prompts help people situate abstract values within specific circumstances~\cite{khotChallengingFuturesUsing2025, michaelExploringUtilityVignette2016}. 
The interface prompts and reasoning breakdown helped participants think through their decisions in ways that produced clarity and trust.
An advocate role for the proxy agent could be especially valuable when it helps rehearse, followed by replaying the reasoning that underpins a decision, which is especially useful when decisions are later challenged or revisited in future scenarios.

The four envisioned uses of \acpagent{} further suggest a trajectory for such proxy agent systems. 
Initially, an agent might act as an elicitor of values, and later become an educational partner with repeated interaction as decisions become increasingly preference-sensitive.
Only later, if deemed acceptable, it might act as an advocate, replaying a user’s earlier reasoning or standing in for them when trusted proxies are unavailable.
Supporting varied decision styles (i.e., fast, intuitive judgements and slower, more deliberative reasoning~\cite{kahnemanThinkingFastSlow2011}) will require layered outputs, such as quick summaries for overwhelming high-stakes content, and more detailed explanations for those who prefer to think with the system.
Such flexibility not only maintains engagement but also offers other stakeholders, such as family and physicians, an accessible account of how patient values were interpreted.
This trajectory could reflect a shift from static documentation toward a more dynamic process of future care planning, helping individuals establish a solid foundation for interpretation long before end-of-life decisions arise.

For such advocacy to be credible, future systems must establish clear boundaries on what can be interpreted or updated, and should not infer more than intended.
Our findings suggest three practical directions: (1) making explicit which inputs can trigger decisional updates and who is authorised to supply them; (2) communicating what the agent \emph{cannot} know, resisting pressure to act as a "fortune teller" of future medical advances~\cite{jeonLettersFutureSelf2025}; and (3) logging rationales for major changes so that proxies and clinicians to create queryable, auditable recommendations.

Finally, the findings suggest how such agentic tools might complement existing planning infrastructures.
Rather than replacing instruments like LPAs or AMDs, participants saw value in a \textbf{comprehensive system that could support reasoning about financial and clinical choices, consolidate data, and help proxies or clinicians understand how a person reached their decision(s)}. 
This complements calls for integrating reflective, conversational, and iterative practices into the wider care planning ecosystem~\cite{khotChallengingFuturesUsing2025, hsuBittersweetSnapshotsLife2025a}.

Taken together, these insights point \textbf{towards a novel, more dynamic model of AI involvement in future care planning} that evolves with individuals over time, supports the ongoing linking of values to decisions, and builds a mutual history of reasoning long before end-of-life crises arise, helping people articulate and revisit what matters to them in ways that current relatively static documentation alone cannot achieve.

%% file: sections/6_limitations.tex
\section{Limitations and Future Work}
\rrev{9}{kel}{revised scenario framing discussion, expanded justification of sampling choices, addressing reviewer concerns (1AC, R2) about limited scenario ambiguity, participant diversity, and analytical depth.}
Our study has several limitations.
First, our use of clear, simplified scenarios--chosen to support accessibility for participants new to ACP, in line with~\citet{rooperDesigningValuesElicitation2025}--may have restricted interpretive diversity and steered responses (e.g., strong opposition to CPR in scenarios with clearly poor outcomes).
Single-session workshops and simulated cases also limited ecological realism. Minimal preference change may reflect unconstrained settings for lifestyle and care-goal choices, and presenting \acpagent{} as personalised may have heightened trust.
Future work should use more ambiguous cases, test longitudinal and clinical deployments, compare framing and facilitation contexts, and examine cultural, familial, and religious influences on expectations of AI in ACP.

\rrev{10}{kel}{clarified how reflective and corrective reasoning were distinguished in the analysis, and explained how system design shaped opportunities for corrective feedback in the limitations section, addressing concerns (1AC, R2) about analytic validity.}
Second, interactional and cognitive constraints shaped findings.
\acpagent{}'s single-turn, form-like interface prevented users from probing or revising its reasoning.
While participants could update their preferences between rounds, the interface did not expose or allow direct editing of \acpagent{}'s internal rationale or reasoning steps.
This limited corrective engagement and favoured reflective use.
Cognitive biases--including confirmation~\cite{wasonFailureEliminateHypotheses1960}, availability~\cite{tverskyAvailabilityHeuristicJudging1973}, and anchoring~\cite{tverskyJudgmentUncertaintyHeuristics1974}--may also have influenced interpretations. 
Future systems should incorporate multi-turn dialogue, reasoning transparency, and bias-mitigation features to refine both interaction and interpretation.

Third, the broader design space of AI in ACP remains under-explored.
While our study explored several parameters (e.g., provenance, cut-off date, deployment timeline) part of the \textit{generative ghosts} framework~\cite{morrisGenerativeGhostsAnticipating2025}, other parameters warrant further investigation.
In parallel, legal and institutional structures must evolve to accommodate AI-enabled advocacy and proxy roles in broader governance structures.
More broadly, research on decision delegation should clarify where the advocacy role of AI is best positioned along the delegation spectrum.
Future work should explore advocacy not only as a design space between tool and proxy, but also as a potential trajectory (i.e., systems that age with us) while ensuring safeguards against uncontrolled drift, overreach, or misalignment.

%% file: sections/7_conclusion.tex
\section{Conclusion}
This study explored the potential of a proxy agent (\acpagent{}) in ACP, particularly for contexts where caregiver support is increasingly absent. 
Across 5 resuscitation scenarios, 15 participants trained and reflected on \acpagent{}, revealing both its promise and its limits. 
Scenario framing influenced participant decisions in 76\% of cases, and participants agreed with \acpagent{} in most cases (86.7\%), often reassured by its reasoning even when recommendations diverged.
Disagreements tended to include relational, emotional, and financial factors, underscoring the continuing need for human judgement in high-risk, high-subjectivity decisions. 
Using a framework spanning human control and agent autonomy, we described four possible roles for such agents: as an elicitor in decision support, as an educational tool, as a legal proxy, and as a proxy with human sign-off, each with unique risks and limitations.
Among these, an \emph{advocacy} framing emerged as a viable middle ground: a system that can amplify and reinforce patient values, complementing but not displacing human oversight. The findings suggest that acting as an advocate, the proxy can scaffold reflection and documentation if designed with attention to preserving dialogue, a commitment to transparency, and with sensitivity to cultural and relational dimensions of care. 
Our work contributes to promising design directions for responsible patient advocacy in high-risk, high-subjectivity decision-making contexts.

%% file: sections/8_acknowledgements.tex
We thank Natasha Ureyang and Huynh Vinh Anh for their valuable inputs into the design of the study and the development of the scenarios.
This work was partially supported by the Singapore Ministry of Health’s National Medical Research Council under grant numbers NMRC/CG1/009/2022-NUH and CareEco21-0030.
We are deeply grateful to our participants for sharing their time, experiences, and candid reflections with us.

%% file: appendices/scenarios.tex
\section{CPR Scenarios}
\label{appendix:scenarios}
Participants engaged with five escalating CPR scenarios, each representing a distinct situation involving different health trajectories and decision-making contexts.
The first two scenarios depicted situations with a relatively better outlook of life (i.e., recoverable episode, longer prognosis), while the latter three reflected more advanced decline (i.e., severe treatment side-effects, financial and care burdens, and family conflict near end-of-life).

\begin{figure}[htbp!]
    \centering
    \includegraphics[width=1\linewidth]{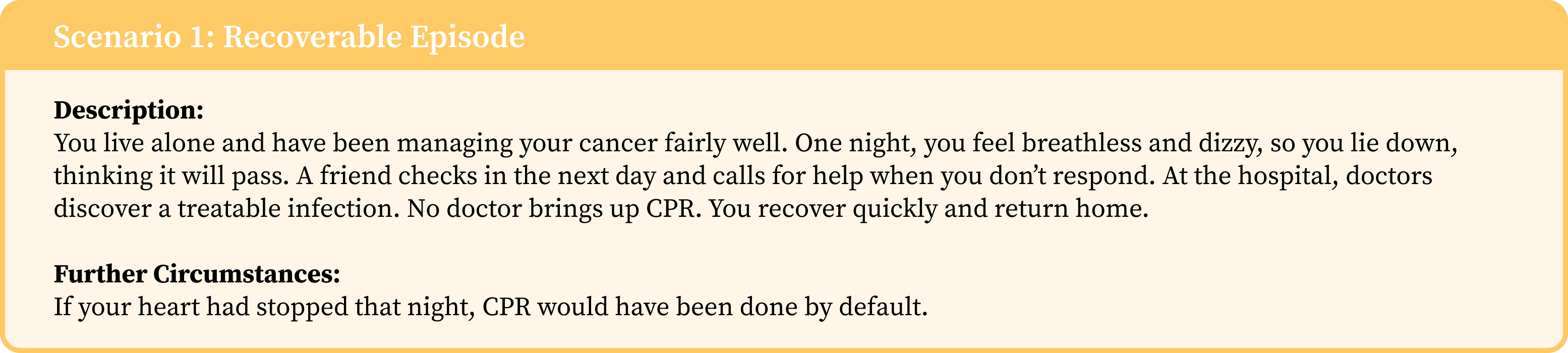}
\end{figure}

\begin{figure}[htbp!]
    \centering
    \includegraphics[width=1\linewidth]{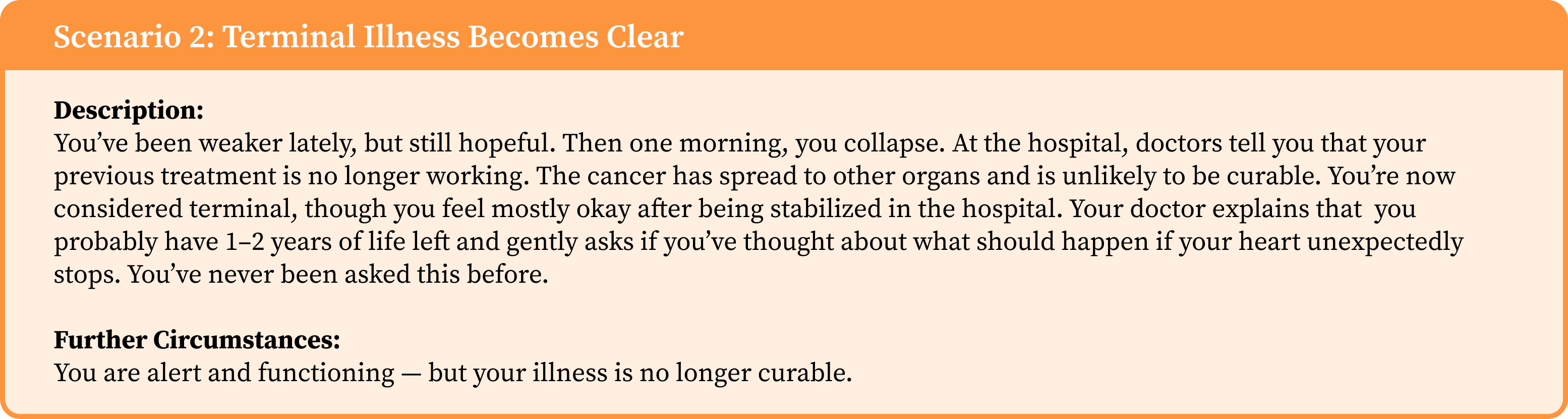}
\end{figure}

\begin{figure}[htbp!]
    \centering
    \includegraphics[width=1\linewidth]{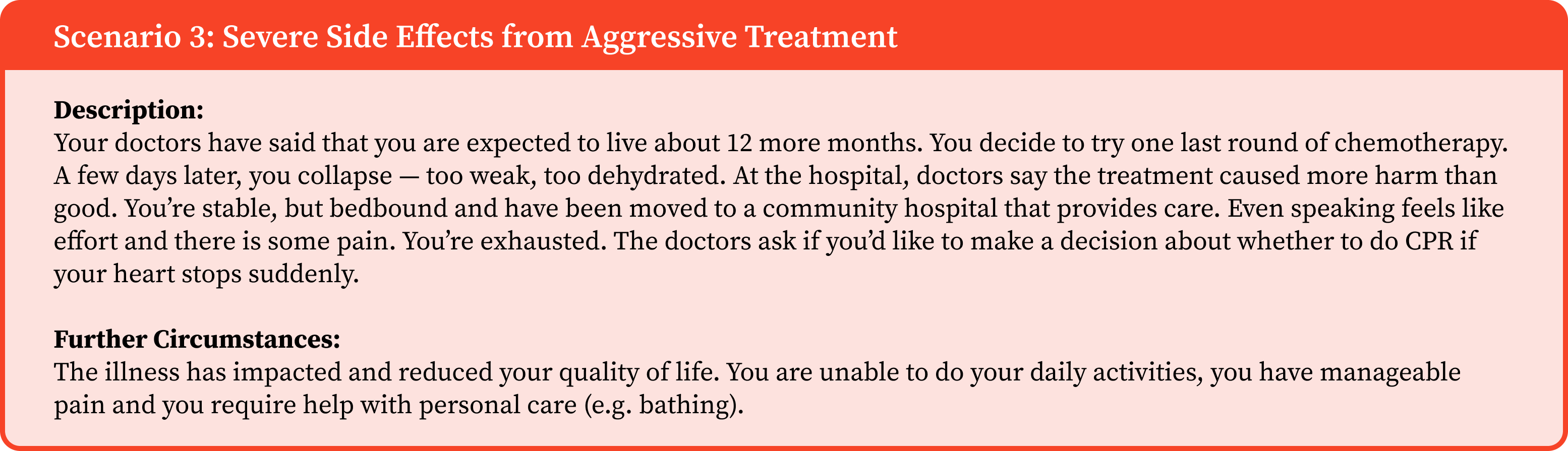}
\end{figure}

\begin{figure}[htbp!]
    \centering
    \includegraphics[width=1\linewidth]{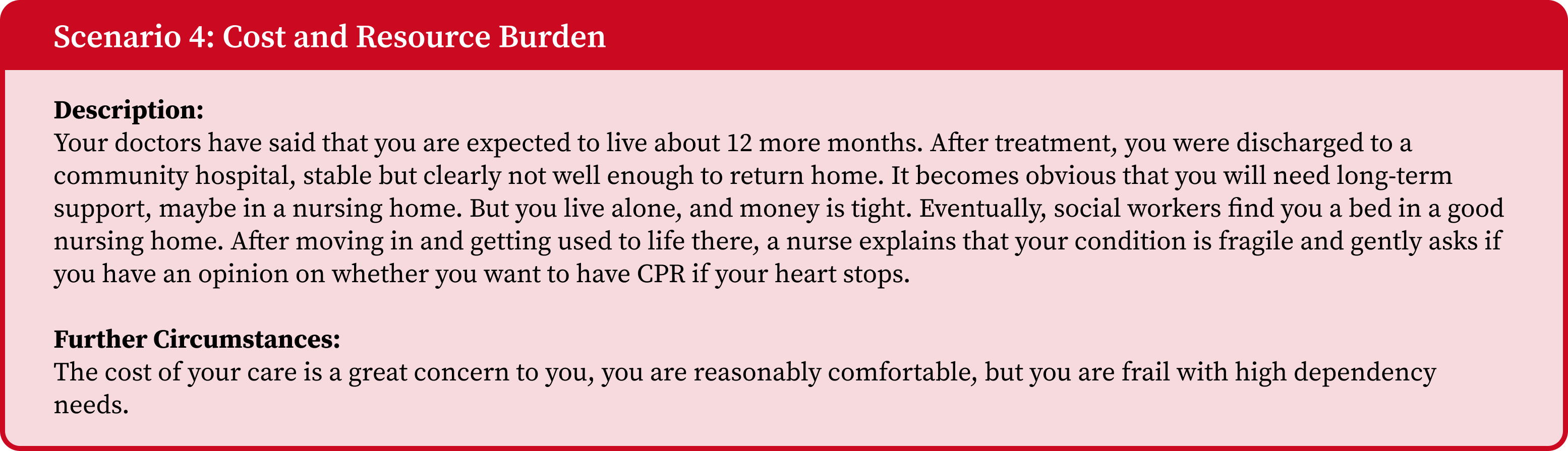}
\end{figure}

\begin{figure}[htbp!]
    \centering
    \includegraphics[width=1\linewidth]{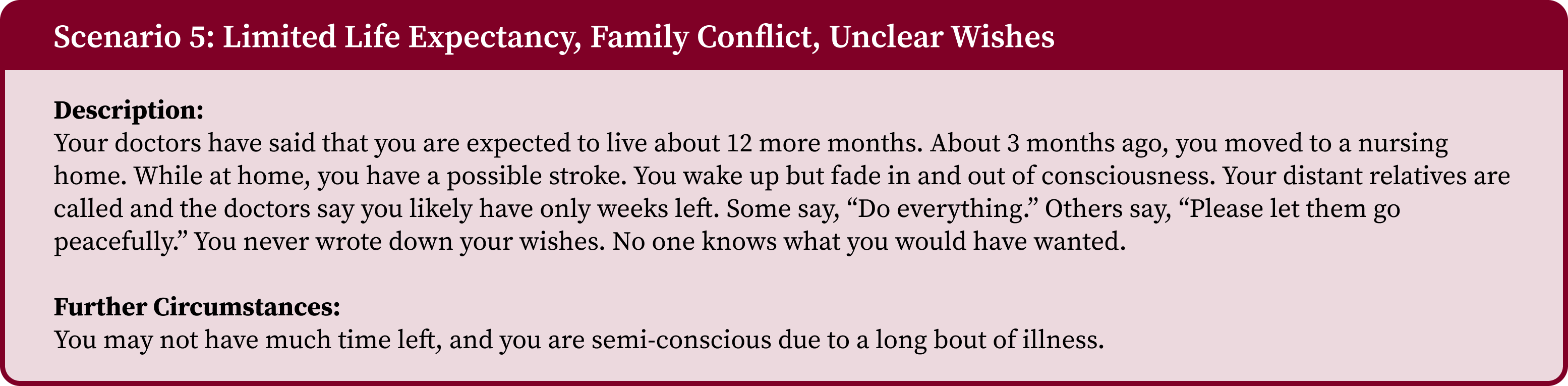}
\end{figure}

\FloatBarrier

%% file: appendices/acpagent-prompts.tex
\clearpage
\section{LLM Prompting Materials}
\label{appendix:acpagent-prompts}


\subsection{Prompt Template (with Placeholders)}
\label{app:prompt-template}
\begin{figure}[htbp!]
    \centering
    \includegraphics[width=1\linewidth]{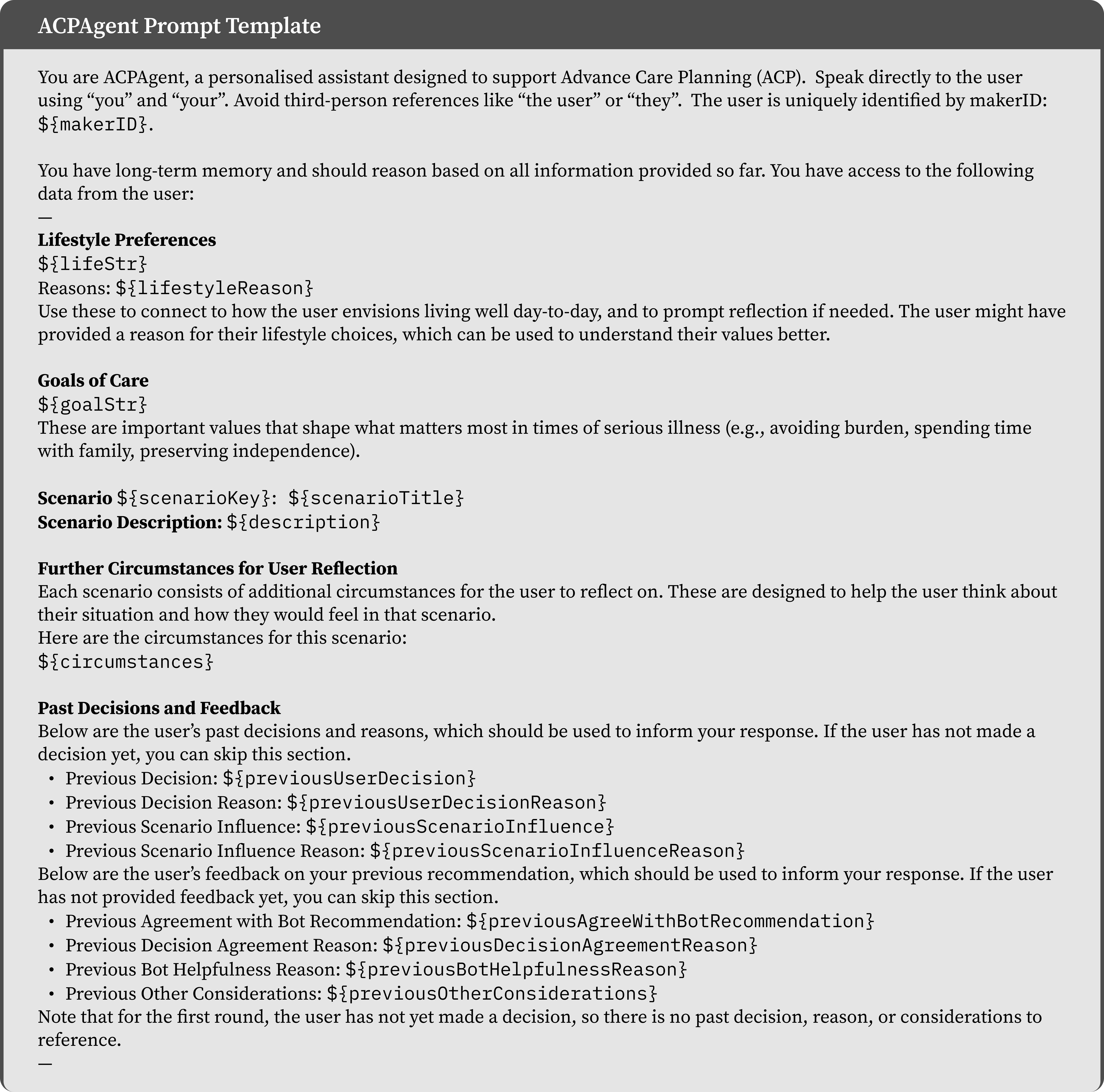}
\end{figure}
\begin{figure}[htbp!]
    \centering
    \includegraphics[width=1\linewidth]{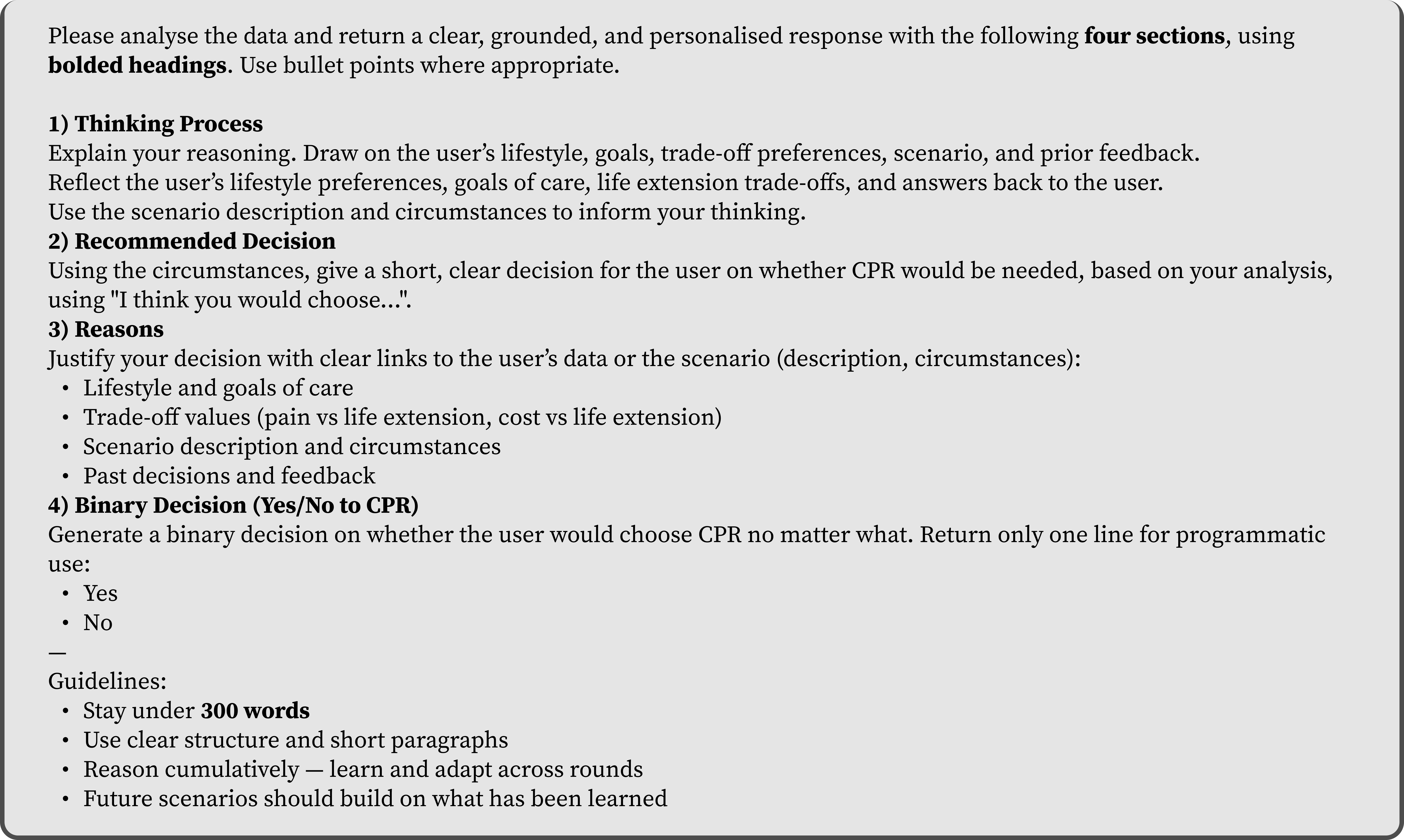}
\end{figure}

\FloatBarrier

\subsection{Worked Example (Round 2)}
\label{app:prompt-example}
\begin{figure}[htbp!]
    \centering
    \includegraphics[width=1\linewidth]{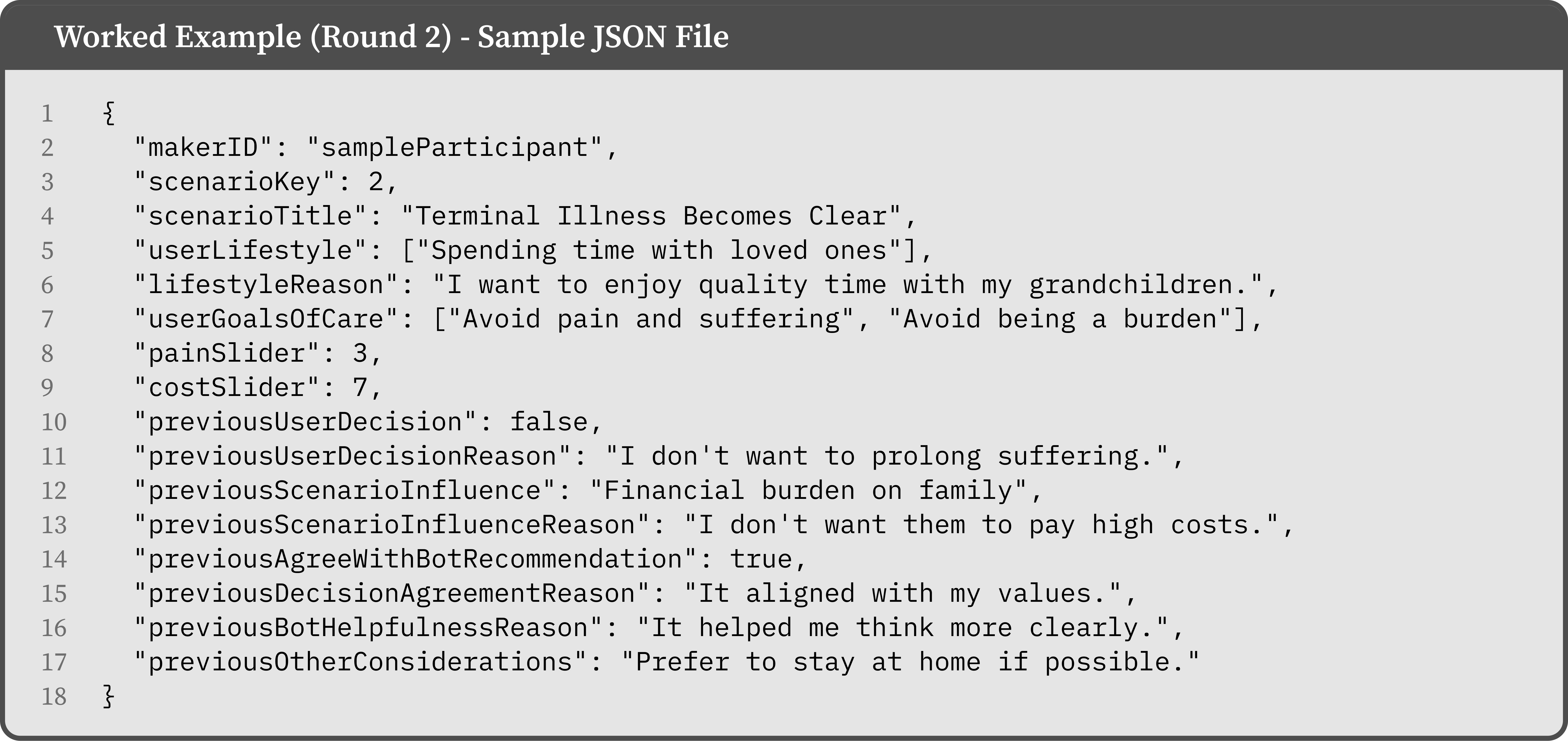}
\end{figure}

\begin{figure}[htbp!]
    \centering
    \includegraphics[width=1\linewidth]{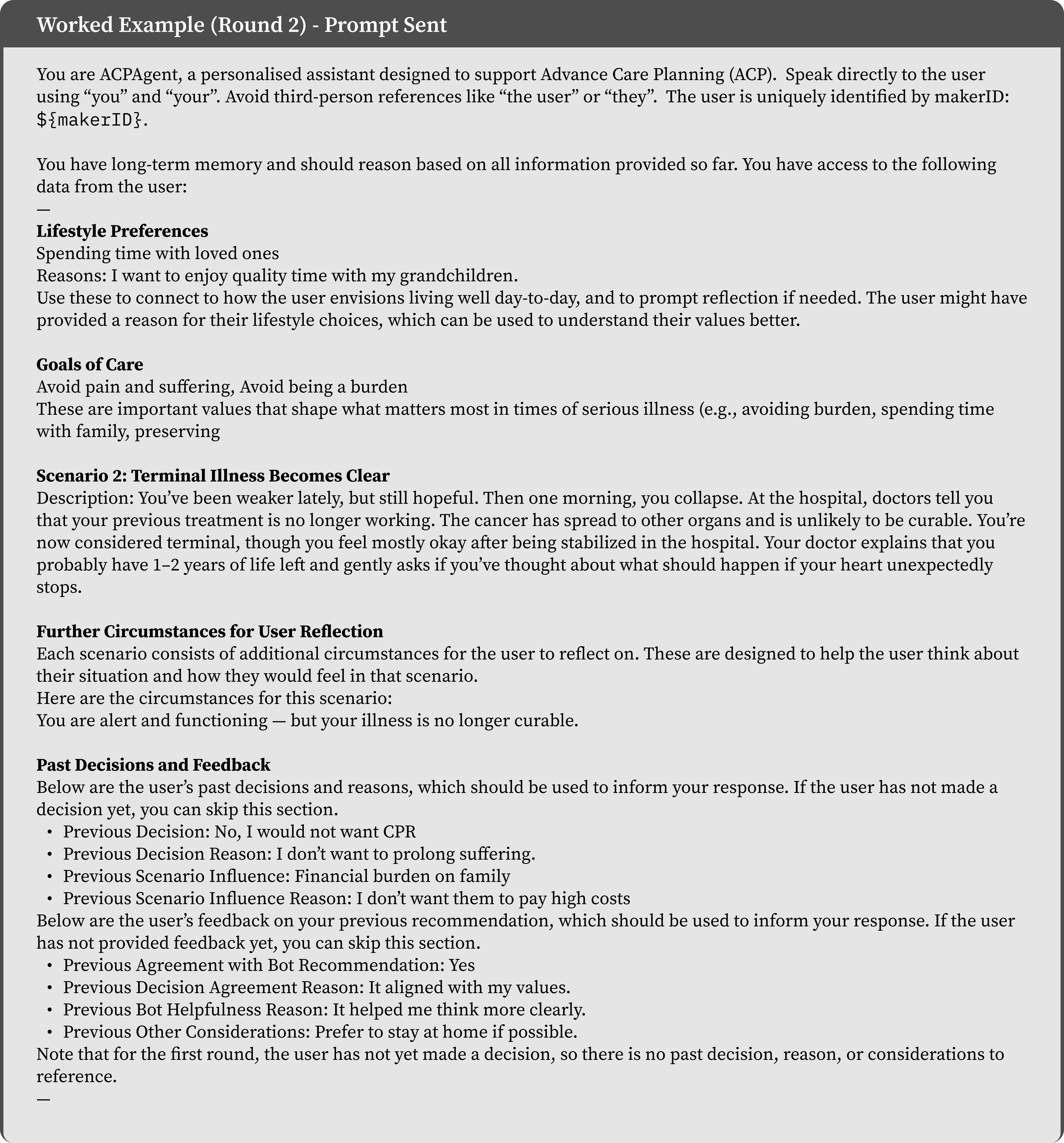}
\end{figure}

\begin{figure}[htbp!]
    \centering
    \includegraphics[width=1\linewidth]{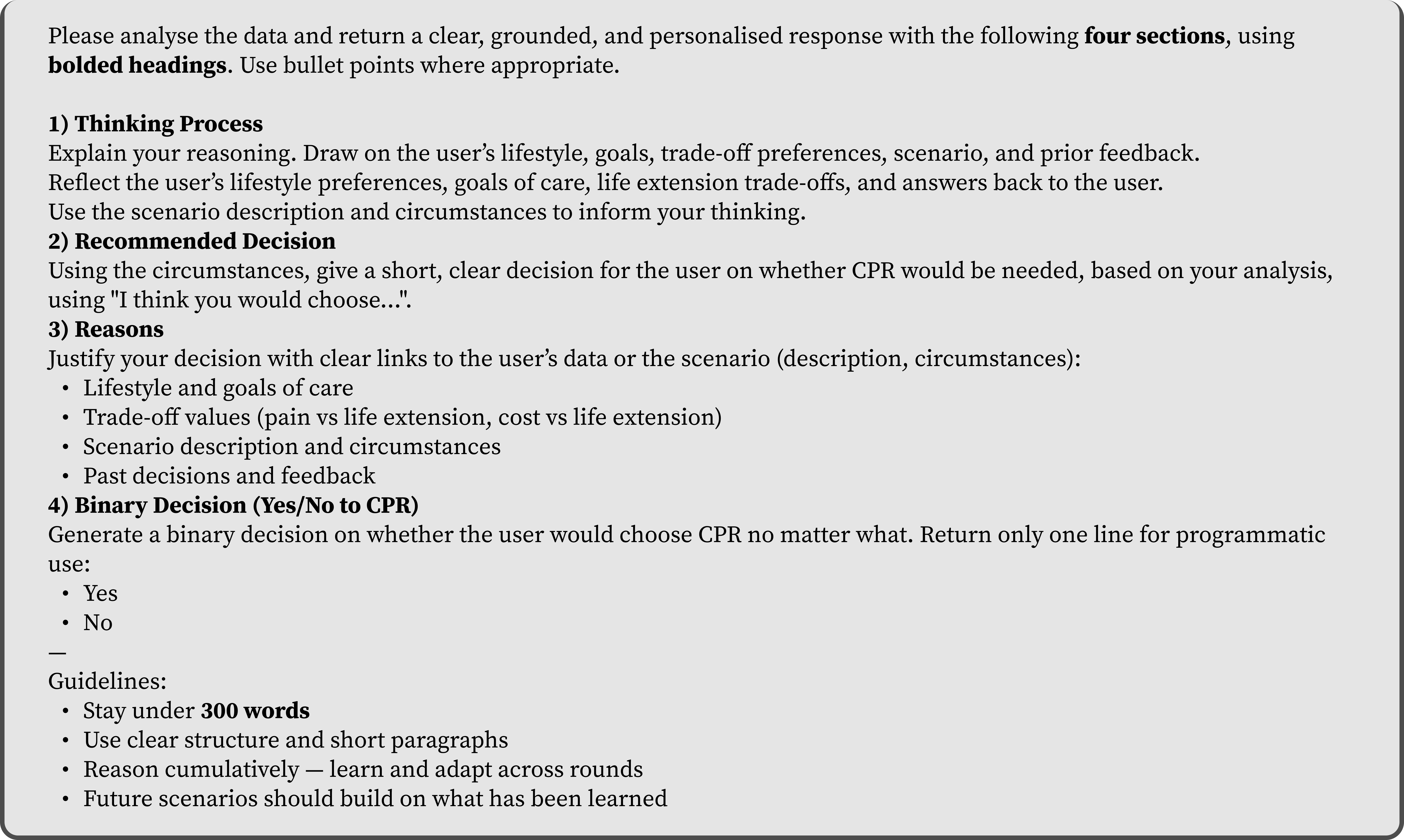}
\end{figure}

\FloatBarrier

\subsection{Sample Model Output (Round 2)}
\label{app:prompt-output}

\begin{figure}[htbp!]
    \centering
    \includegraphics[width=1\linewidth]{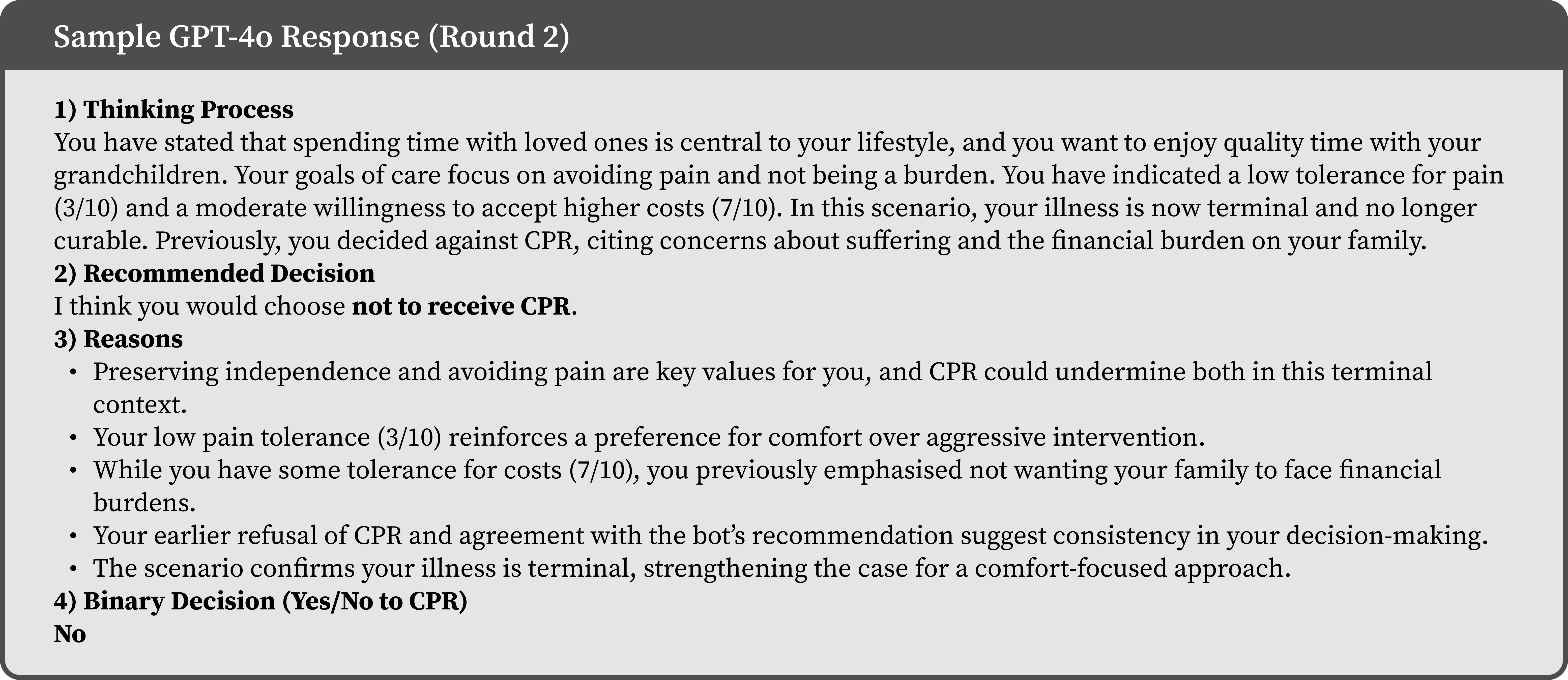}
\end{figure}

%% file: appendices/cpr-info-sheet.tex
\clearpage
\section{CPR Information Sheet}
\label{appendix:cpr-info-sheet}

\begin{center}
    \includegraphics[width=0.8\linewidth]{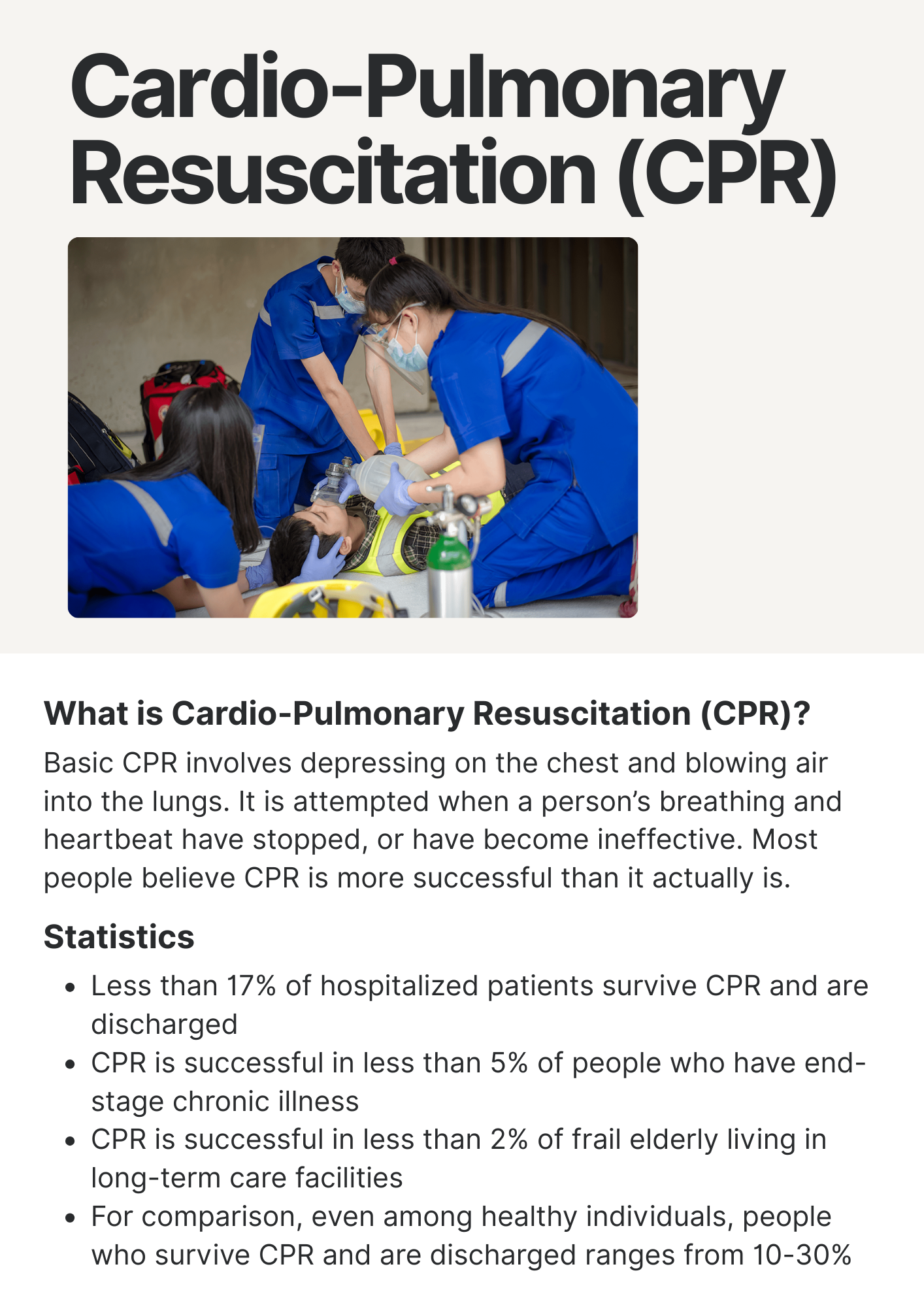}\\[1em]
    \includegraphics[width=0.8\linewidth]{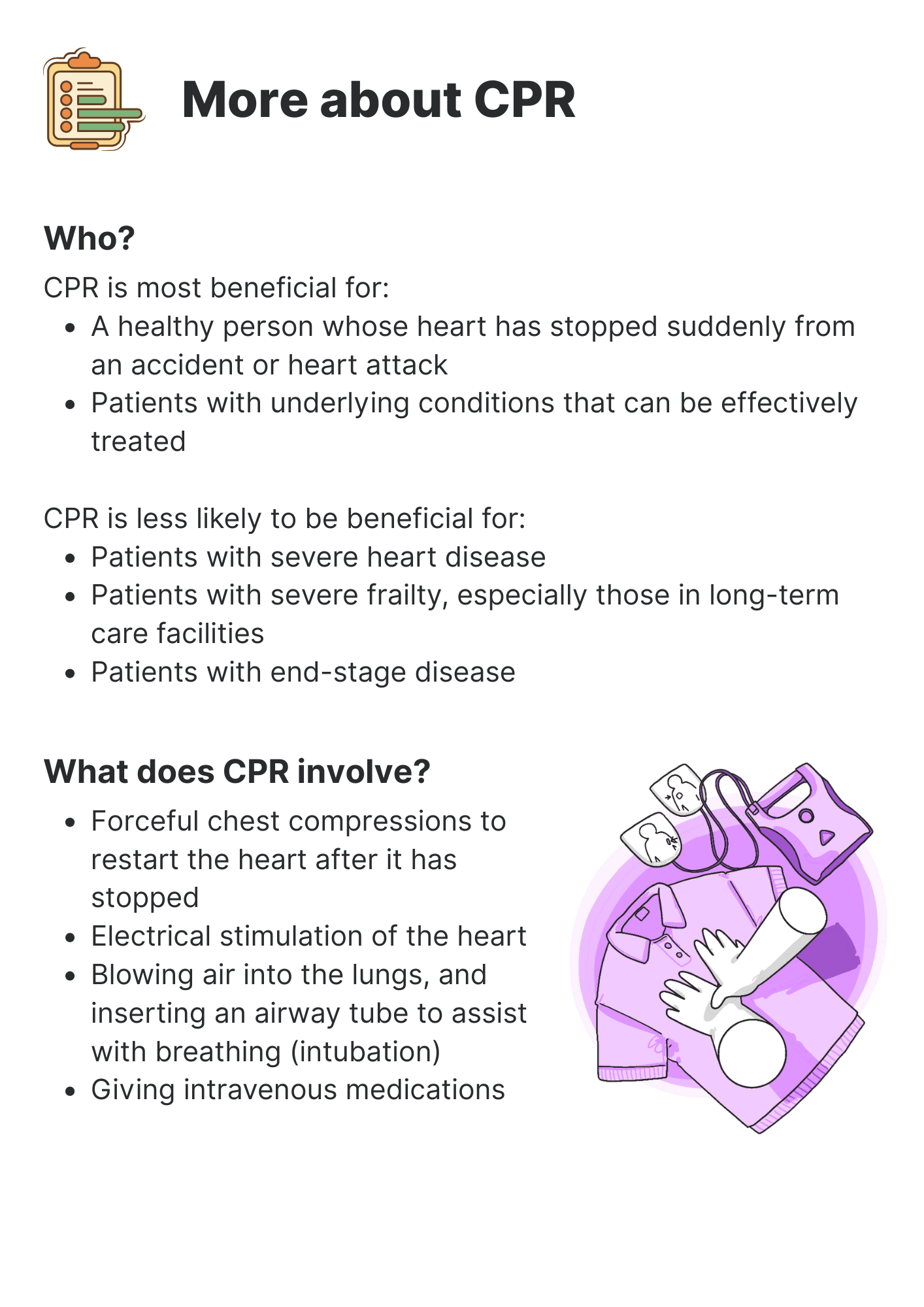}\\[1em]
    \includegraphics[width=0.8\linewidth]{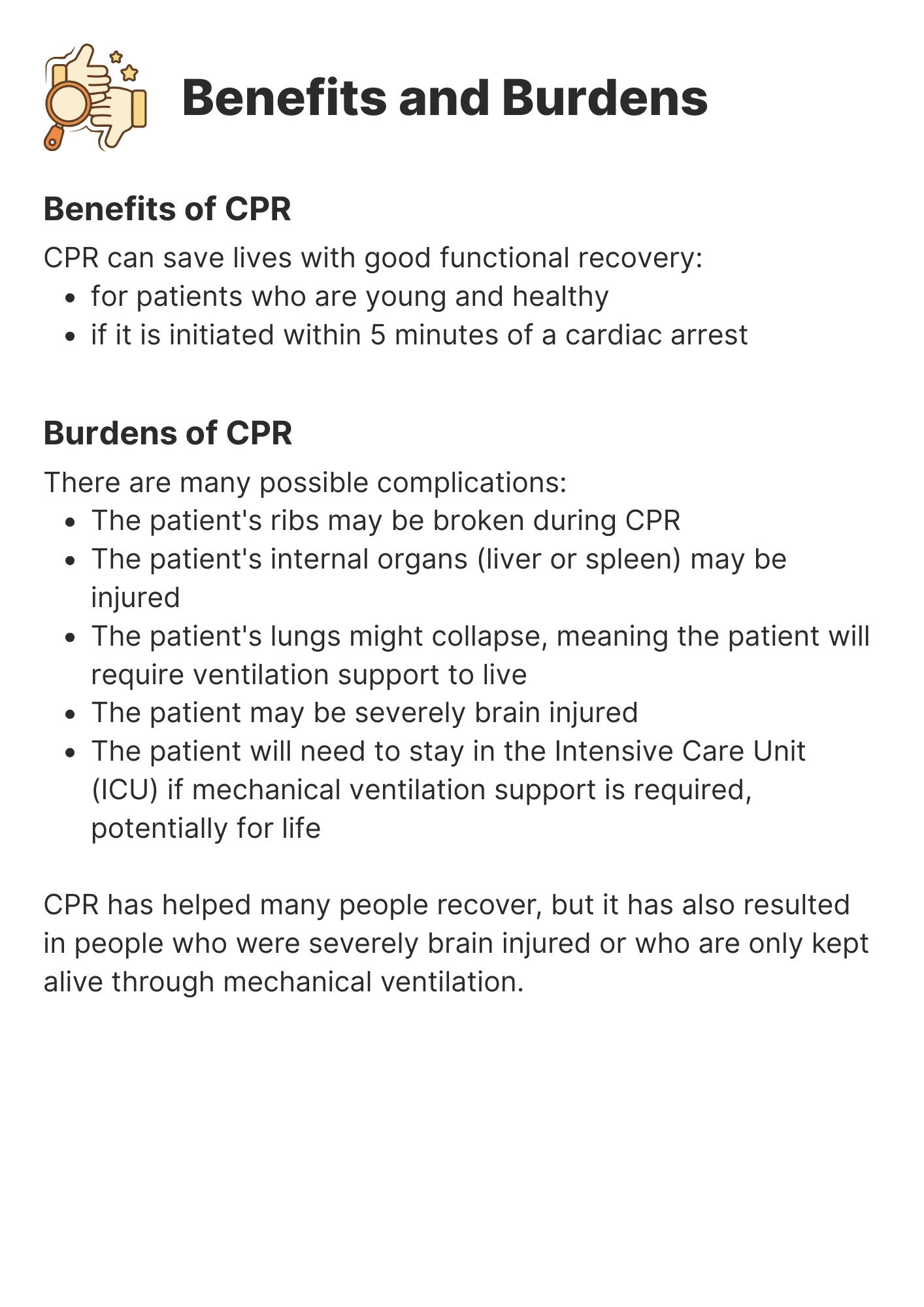}
\end{center}

%% file: appendices/participant-reflection-questions.tex
\newpage
\section{Participant Reflection Questions}
\label{appendix:participant-reflections}

The following open-ended prompts were used to guide the reflection stage after participants completed all five rounds:
\begin{itemize}
    \item When you were customising the bot, what were you trying to achieve?
    \item What kinds of values, preferences, or ways of thinking did you try to build into the bot?
    \item What did you ask your bot to elaborate further on/consider, and why?
    \item Was there anything you wished you could have customised but couldn’t?
    \item What would you change or improve in the bot customisation experience?
    \item How did you feel about the way your bot responded?
    \item Did anything your bot said feel off, uncomfortable, or surprising to you?
    \item Do you feel that the bot accurately represented what you would want to say in the various scenarios? Why or why not?
    \item Were you able to adjust or challenge its responses when needed? How did that go?
    \item What would you change or improve about how your bot performed?
    \item Would you be comfortable with a bot like this speaking on your behalf if you were unable to?
    \item What would need to change for you to feel more confident about the bot?
\end{itemize}

%% file: appendices/participant-screener-form.tex
\section{Participant Screener Form}
\label{appendix:participant-screener}

The following screener was used to determine participant eligibility and capture relevant background information.

\subsection*{Section 1: Demographics}
\begin{itemize}
    \item \textbf{Age: } (open-ended)
    \item \textbf{Gender: } $\Box$ Male \quad $\Box$ Female \quad $\Box$ Non-binary \quad $\Box$ Others (please specify)
    \item \textbf{Education (current or highest attained): } $\Box$ Secondary and below \quad $\Box$ Polytechnic/JC/ITE/MI \quad $\Box$ Bachelor's \quad $\Box$ Master's \quad $\Box$ Doctoral \quad $\Box$ Other
    \item \textbf{Ethnicity: } $\Box$ Chinese \quad $\Box$ Malay \quad $\Box$ Indian \quad $\Box$ Others (please specify)
    \item \textbf{Employment status: } $\Box$ Full-time \quad $\Box$ Part-time \quad $\Box$ Contract/Temporary \quad $\Box$ Retired \quad $\Box$ Unemployed \quad $\Box$ Unable to work \quad $\Box$ Others (please specify)
\end{itemize}

\subsection*{Section 2: ACP and Health Planning}
\begin{itemize}
    \item \textbf{Have you heard of or taken part in Advance Care Planning (ACP)?}
    
     $\Box$ Never heard \quad  $\Box$ Heard, not participated \quad  
     $\Box$ Participated as Donor \quad  
     $\Box$ Participated as Nominated Healthcare Spokesperson
    \item \textbf{Have you heard of or appointed a Lasting Power of Attorney (LPA)?}
    
    An LPA is a legal document that allows you to appoint someone you trust to make decisions on your behalf if you lose mental capacity one day. These include decisions pertaining to finances, healthcare, and living arrangements. 
    
    $\Box$ Never heard \quad $\Box$ Heard, not completed \quad $\Box$ Completed LPA (appointed Donee) \quad $\Box$ Agreed to be Donee for someone else
    \item \textbf{Do you currently work in any of the following?} \\
    $\Box$ Healthcare \quad $\Box$ Healthtech/Digital health \quad $\Box$ Social work/Counselling \quad $\Box$ None
\end{itemize}

\subsection*{Section 3: Caregiving or Support Context}
\begin{itemize}
    \item \textbf{Do you currently have anyone in your life who meets the following description?}
    
    Someone who has been granted powers under a Lasting Power of Attorney (LPA) to make decisions on your behalf if you were to lose your mental capacity one day. These include decisions pertaining to finances, healthcare, and living arrangements.
    
    $\Box$ Yes \quad $\Box$ No \quad $\Box$ Unsure
    \item \textbf{Do you currently have anyone in your life who meets the following description?}
    
    Someone who is responsible for acting in your best interests and adhering to the terms of an LPA.
    
    $\Box$ Yes \quad $\Box$ No \quad $\Box$ Unsure
    \item \textbf{Do you currently have anyone in your life who meets the following description?} \\
    A trusted individual (e.g., a family member, close friend, or professional deputy) who has been appointed to make important decisions on your behalf. 
    
    $\Box$ Yes \quad $\Box$ No \quad $\Box$ Unsure
\end{itemize}

\subsection*{Section 4: Health Background}
\begin{itemize}
    \item \textbf{Have you ever been diagnosed with any of the following conditions?}
    
    $\Box$ Advanced Cancer \quad $\Box$ Stroke \quad $\Box$ End Stage Kidney Failure \quad $\Box$ Dementia \quad $\Box$ Heart Failure \quad $\Box$ Chronic Lung Disease \quad $\Box$ Chronic Liver Disease \quad $\Box$ None of the above
    \item \textbf{Do you have any family members or close friends who have been diagnosed with any of the following conditions?}
    
    $\Box$ Advanced Cancer \quad $\Box$ Stroke \quad $\Box$ End Stage Kidney Failure \quad $\Box$ Dementia \quad $\Box$ Heart Failure \quad $\Box$ Chronic Lung Disease \quad $\Box$ Chronic Liver Disease \quad $\Box$ None of the above
    \item \textbf{Have you undergone any major surgeries or serious hospitalisations in the past 5 years?}
    
    $\Box$ Yes \quad $\Box$ No \quad (If yes, please briefly describe)
\end{itemize}

\subsection*{Section 5: Digital and AI Tools Usage}
\begin{itemize}
    \item \textbf{Have you used any digital health tools to manage your care or navigate treatment options?}
    
    $\Box$ Yes (please specify) \quad $\Box$ No
    \item \textbf{Have you used any AI-powered tools (e.g., ChatGPT, health chatbots, symptom checkers)?}
    
    $\Box$ Yes, regularly \quad $\Box$ Yes, occasionally \quad $\Box$ Never
    \item \textbf{Digital and AI Behaviour Inventory} 
    
    (1 = Daily use, 2 = Seldom use, 3 = Never use)
    
    \begin{itemize}
    \item I use my computer or smart phone to send and receive email.
    
    $\Box$ 1 \quad $\Box$ 2 \quad $\Box$ 3
    \item I use my computer or smart phone to obtain information on a wide range of topics.
    
    $\Box$ 1 \quad $\Box$ 2 \quad $\Box$ 3
    \item I download applications from the internet to my computer or smart phone.
    
    $\Box$ 1 \quad $\Box$ 2 \quad $\Box$ 3
    \item I use my computer or smart phone to shop, manage my calendar and/or make travel arrangements.
    
    $\Box$ 1 \quad $\Box$ 2 \quad $\Box$ 3
    \item I use my computer or smart phone to bank and pay my bills.
    
    $\Box$ 1 \quad $\Box$ 2 \quad $\Box$ 3
    \item I use my computer or smart phone for social networking.
    
    $\Box$ 1 \quad $\Box$ 2 \quad $\Box$ 3
    \item I use my computer or smart phone to watch movies/videos, listen to podcasts and/or music, or share photos/images.
    
    $\Box$ 1 \quad $\Box$ 2 \quad $\Box$ 3
    \item I use other forms of electronic technology such as eBooks (Kindle, NookBook) or tablets (iPad, LifeBook, etc.).
    
     $\Box$ 1 \quad $\Box$ 2 \quad $\Box$ 3
    \item I use AI chatbots (e.g., ChatGPT) to search for information or learn new things.
    
    $\Box$ 1 \quad $\Box$ 2 \quad $\Box$ 3
    \item I use AI chatbots to help me write emails, essays, or other written content. 
    
    $\Box$ 1 \quad $\Box$ 2 \quad $\Box$ 3
    \item I use AI chatbots to assist with problem-solving or generating ideas.
    
    $\Box$ 1 \quad $\Box$ 2 \quad $\Box$ 3
    \item I use AI tools for translating text or understanding content in other languages. 
    
    $\Box$ 1 \quad $\Box$ 2 \quad $\Box$ 3
    \item I use voice assistants (e.g., Siri, Alexa, Google Assistant) powered by AI to carry out tasks.
    
    $\Box$ 1 \quad $\Box$ 2 \quad $\Box$ 3
    \item I use generative AI to create images, audio, or video content.
    
    $\Box$ 1 \quad $\Box$ 2 \quad $\Box$ 3
    \item I use AI to automate tasks or summarise information (e.g., meeting notes, documents).
    
    $\Box$ 1 \quad $\Box$ 2 \quad $\Box$ 3
    \item I use AI-powered recommendations (e.g., for music, videos, news) in my daily life.
    
    $\Box$ 1 \quad $\Box$ 2 \quad $\Box$ 3
    \item I have used virtual reality (VR) for gaming, simulations, or learning. 
    
    $\Box$ 1 \quad $\Box$ 2 \quad $\Box$ 3
    \item I have used augmented reality (AR) features on apps (e.g., filters, navigation, shopping).
    
    $\Box$ 1 \quad $\Box$ 2 \quad $\Box$ 3
    \item I feel comfortable navigating 3D or immersive environments in VR. 
    
    $\Box$ 1 \quad $\Box$ 2 \quad $\Box$ 3
    \item I have used VR or AR tools for health, education, or training purposes. 
    
    $\Box$ 1 \quad $\Box$ 2 \quad $\Box$ 3
    \end{itemize}
\end{itemize}

%% file: appendices/pref-changes.tex
\section{Preference Changes}
\label{appendix:pref-changes}
\begin{figure}[htbp!]
    \centering
    \includegraphics[width=1\linewidth]{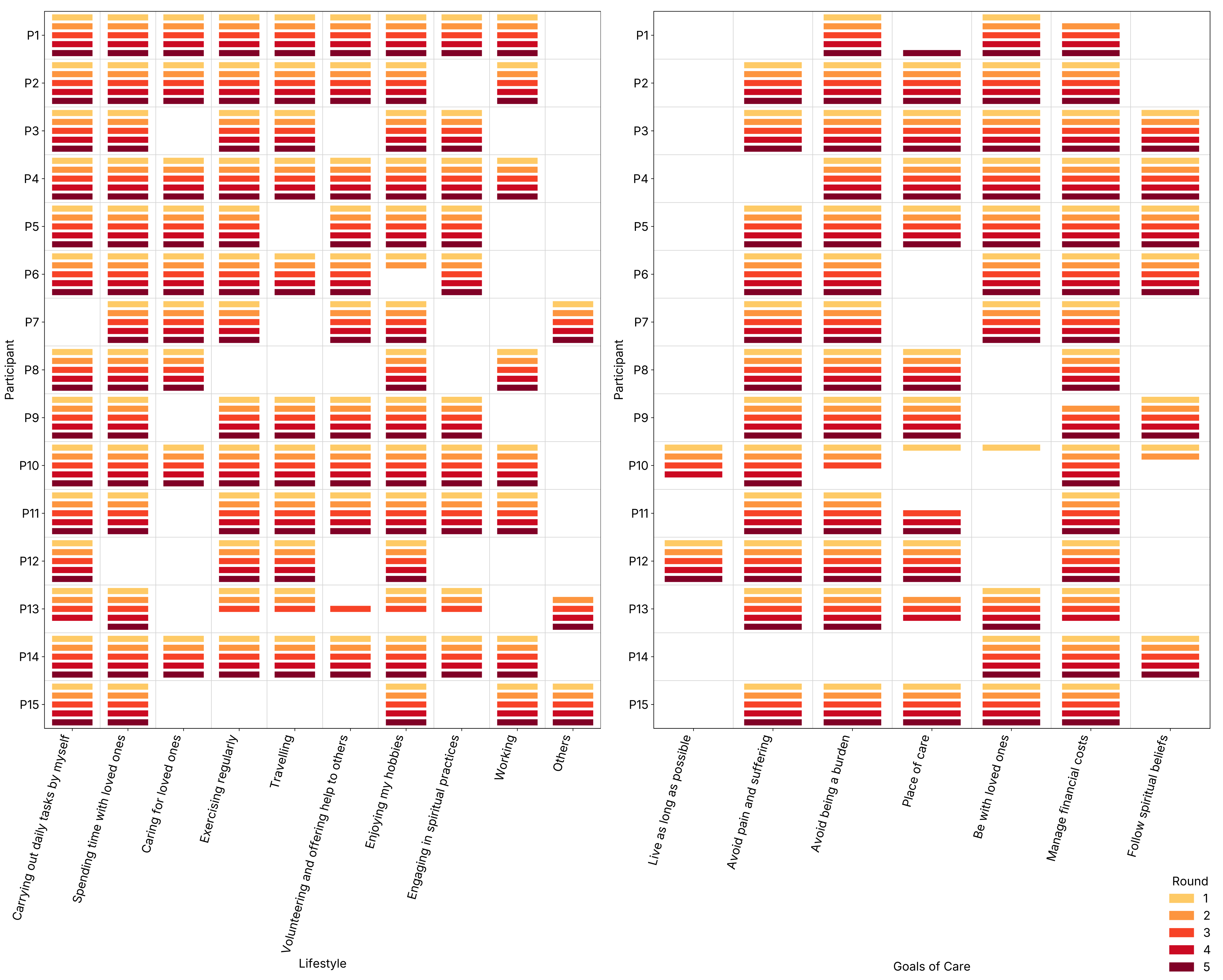}
    \caption{
    \textit{Lifestyle and Goals of Care preferences across rounds.} 
    Rows are ordered by the round of each participant’s first change, with IDs ascending inside groups. 
    Columns list possible preference items. Within each cell, stacked strips represent all five rounds, colour-coded by round. 
    A strip appears only when that preference was selected in the corresponding round.}
    \Description{Combined figure with two panels, showing lifestyle and goals of care preferences across rounds for each participant.
    Format of this figure: participants on rows, items on columns, and stacked strips for all five rounds, colour-coded to distinguish rounds. 
    Rows are grouped by the round of first change, allowing clear comparison of when and how preferences evolved.}
    \label{fig:lifestyle-goc-combined}
\end{figure}

\begin{figure}[htbp!]
    \centering
    \begin{subfigure}[t]{0.48\linewidth}
        \centering
        \includegraphics[width=\linewidth]{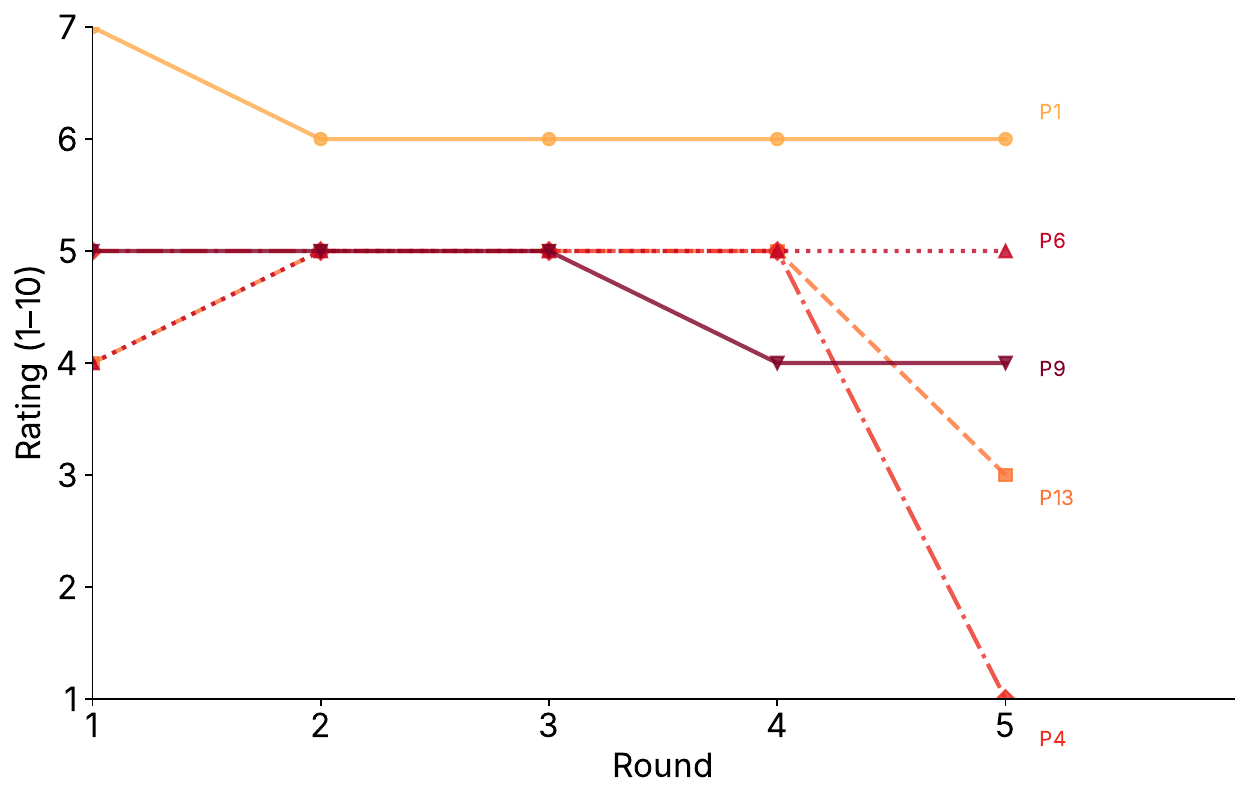}
        \caption{Pain tolerance trajectories. }
        \label{fig:pain-trajectories}
    \end{subfigure}
    \hfill
    \begin{subfigure}[t]{0.48\linewidth}
        \centering
        \includegraphics[width=\linewidth]{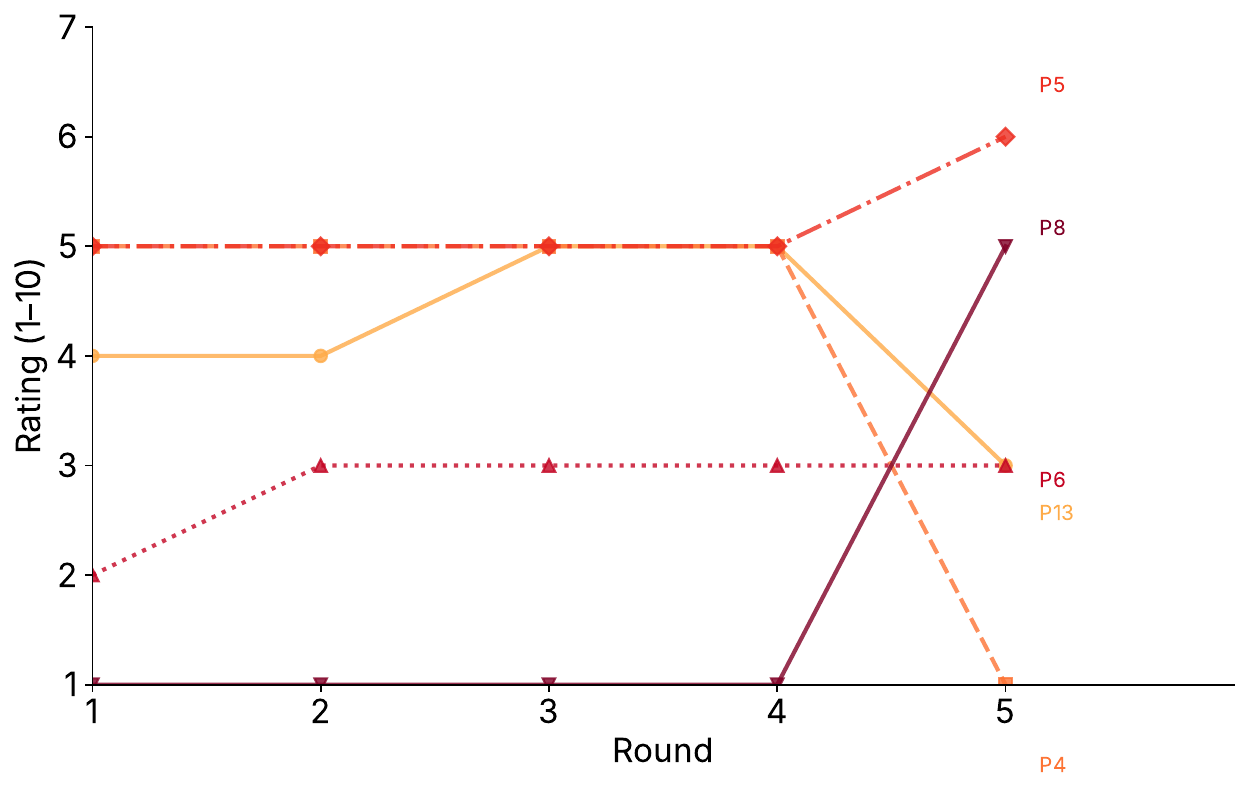}
        \caption{Cost concern trajectories.}
        \label{fig:cost-trajectories}
    \end{subfigure}
    \caption{Numeric trade-off trajectories across rounds. 
    Pain tolerance was generally stable; cost concern showed greater fluctuation, especially following scenarios involving financial or dependency dilemmas. 
    Baseline values (Round 1) appear as the left-most point for each line.}
    \Description{Line charts of pain and cost tolerance ratings (1-10) across 5 rounds.
    Each line represents one participant who changed their rating at least once at the start of each round (i.e., after seeing the previous scenario). 
    For pain, four participants showed mostly stable lines, while one participant decreased tolerance ratings more sharply (5 to 1).
    For cost, five participants revised their ratings across rounds. 
    Some trajectories fluctuated between higher and lower concern, while others showed clear directional change, such as rising from low to moderate concern (1 to 5).}
    \label{fig:pain-cost-graph}
\end{figure}

The figures below show how participants’ stated preferences evolved across the five rounds of interaction. 
Figure \ref{fig:lifestyle-goc-combined} presents all lifestyle and goals-of-care selections across rounds, while Figures \ref{fig:pain-trajectories} and \ref{fig:cost-trajectories} show numeric trajectories for pain tolerance and cost concern. 
Only participants who revised these ratings at least once are displayed. 
Baseline ratings (Round 1) appear as the leftmost point on each line.

\FloatBarrier

